\documentstyle[12pt,aaspp4,psfig]{article}
\begin{document}

\title{The Distribution of Pressures in a Supernova-Driven
Interstellar Medium. I. Magnetized Medium}

\author{Mordecai-Mark Mac Low\altaffilmark{1}} 
\author{Dinshaw S. Balsara\altaffilmark{2}} 
\author{Jongsoo Kim\altaffilmark{2,3}}
\author{Miguel A. Avillez\altaffilmark{1,4}}
\altaffiltext{1}{Department of Astrophysics, American Museum of
  Natural History, Central Park West at 79th Street, New York, NY,
  10024-5192, USA; E-mail: mordecai@amnh.org}
\altaffiltext{2}{Department of Physics, Notre Dame University, 225
  Nieuwland Science Hall, Notre Dame, IN 46556-5670, USA, E-mail:
  dbalsara@nd.edu} 
\altaffiltext{3}{Korea Astronomy Observatory, 61-1, Hwaam-Dong,
  Yusong-Ku, Taejon 305-348, Korea, E-mail: jskim@kao.re.kr} 
\altaffiltext{4}{Department of Mathematics, University of \'Evora,
  R. Rom\~ao Ramalho 59, P-7000 \'Evora, Portugal, E-mail:
  mavillez@galaxy.lca.uevora.pt}

\begin{abstract}
  Observations have suggested substantial departures from pressure
  equilibrium in the interstellar medium (ISM) in the plane of the
  Galaxy, even on scales under 50~pc.  Nevertheless, multi-phase
  models of the ISM assume at least locally isobaric gas.  The
  pressure then determines the density reached by gas cooling to
  stable thermal equilibrium.  We use numerical models of the
  magnetized ISM to examine the consequences of supernova driving for
  interstellar pressures.  In this paper we examine a (200~pc)$^3$
  periodic domain threaded by magnetic fields.  Individual parcels of
  gas at different pressures reach widely varying points on the
  thermal equilibrium curve: no unique set of phases is found, but
  rather a dynamically-determined continuum of densities and
  temperatures.  A substantial fraction of the gas remains entirely
  out of thermal equilibrium. Our results appear consistent with
  observations of interstellar pressures.  They also suggest that the
  high pressures observed in molecular clouds may be due to ram
  pressures in addition to gravitational forces.  Much of the gas in
  our model lies far from equipartition between thermal and magnetic
  pressures, with ratios ranging from 0.1 to $10^4$ and ratios of
  uniform to fluctuating magnetic field of 0.5--1.  Our models show
  broad pressure probability distribution functions with log-normal
  functional forms produced by both shocks and rarefaction waves,
  rather than power-law distributions produced by isolated supernova
  remnants. The width of the distribution can be described
  quantitatively by a formula derived from the work of Padoan,
  Nordlund, \& Jones (1997).
\end{abstract}


\clearpage

\section{Introduction}

Theoretical models of the interstellar medium (ISM) have generally
followed Spitzer (1956) in assuming that the interstellar gas is in
pressure equilibrium.  Field, Goldsmith \& Habing (1969; hereafter
FGH) demonstrated that the form of the interstellar cooling and
heating curves for temperatures below about $10^4$~K allowed a range
of pressures in which two isobaric phases could exist in stable
thermal equilibrium.  Although the details of this model turned out to
be incorrect, due to their assumption of a cosmic ray flux much higher
than subsequently observed, Wolfire et al.\ (1995) demonstrated that
photoelectric heating of polycyclic aromatic hydrocarbons or other
very small dust grains could recover a two-phase model for the neutral
ISM.  McKee \& Ostriker (1977; hereafter MO77) used a similar framework
for the cold ISM, but incorporated the suggestion by Cox \& Smith
(1974) that supernovae (SNe) would create large regions of hot gas
with $T\sim 10^6$~K, to produce a model of a three-phase medium.  They
were the first to relax the assumption of global pressure equilibrium,
noting that SN remnants (SNRs) would have varying pressures.  They
assumed only local pressure equilibrium at the surfaces of clouds in
order to determine conditions there.

Measurements of the ISM pressure were performed using {\em Copernicus}
observations of ultraviolet (UV) absorption lines from excited states
of C~{\sc i} by Jenkins \& Shaya (1979) and Jenkins, Jura \&
Loewenstein (1983).  They found greater than an order of magnitude
variation in pressures in the cold gas traced by the C~{\sc i}, with
most of the gas having pressures $P/k < 10^4$~K~cm$^{-3}$, but a small
fraction reaching pressures $P/k > 10^5$~K~cm$^{-3}$.  These findings
have recently been confirmed and extended using {\em Space Telescope
  Imaging Spectrograph} (STIS) observations by Jenkins \& Tripp
(2001).  Bowyer et al.\ (1995) \& Bergh\"ofer et al.\ (1998) compared
average pressures derived from observations with the {\em Extreme UV
  Explorer} of shadows cast by clouds of neutral hydrogen within the
Local Bubble with the pressure of the warm cloud surrounding the Sun
and found a difference of as much as a factor of 25.  McKee (1996)
argued that these estimates were too extreme, but nevertheless
concluded that the observations showed a pressure variation of at
least a factor of five.

In this series of papers we show that dynamical models of a SN-driven
interstellar medium do not produce an isobaric medium controlled by
thermal instability, but rather a medium with a broad pressure
distribution controlled by the dynamics of turbulence, with
rarefaction waves being as important as shocks in setting local
pressures.  Although the cooling curve may determine local behavior,
the pressure of each parcel is set dynamically by flows primarily
driven by distant SNe.  The theory of the probability distribution
function (PDF) of compressible turbulence described by Padoan,
Nordlund \& Jones (1997; hereafter PNJ97) seems able to describe the
pressure PDFs we find here.

Gazol et al.\ (2001) suggest that turbulence driven by ionization
heating results in nearly half of the gas lying in thermally unstable
regions, in agreement with the observations of Heiles (2001).  Earlier
dynamical models of the ISM have been reviewed by Mac Low (2000).
Models including SN driving were first done in two dimensions by Rosen
\& Bregman (1995), and have since been done by a number of other
groups. V\'azquez-Semadeni, Passot \& Pouquet (1995) briefly discussed the
pressure structure in a two-dimensional hydrodynamical model including
only photoionization heating.  In three dimensions, Avillez (2000)
used an adaptive mesh refinement code to compute the evolution of a
$1\times 1\times 20$~kpc vertical section of a galactic disk, while
Korpi et al.\ (1999) used a single-grid magnetohydrodynamical (MHD)
code to model an SN-driven galactic dynamo.
Slyz et al.\ (2005) used a single-grid hydrodynamical code to compute
supernovae in a periodic box including self-gravity and a
star-formation recipe. Both of these papers 
briefly discuss pressure
structure as well. All of these computations show that the
interactions between SNRs drive turbulent flows throughout the ISM,
and that radiative cooling of compressed regions produces cold, dense
clouds with relatively short lifetimes.  

We here study in detail the pressure distribution in MHD models of a
region in the plane of the SN-driven ISM, while in a subsequent paper
we will study the pressure distribution in a stratified ISM.  We use
the Riemann framework of Balsara (2000) in single-grid mode on a
200~pc cube with periodic boundary conditions on all sides, including
both radiative cooling and heating, as well as isolated SNs, 
though neither external or self-gravity.

In \S~\ref{sec:models} we describe the numerical model.  We then give
three different analytic approaches to the question of the pressure
distribution in the SN-driven ISM in \S~\ref{sec:an-th}, derived from
the work of MO77, PNJ97, and Passot \& V\'azquez-Semadeni (1998;
hereafter PV98).  In \S~\ref{sec:numerics} we use our numerical
results to demonstrate that the SN-driven ISM is far from isobaric,
with order of magnitude pressure variations, but that the distribution
of pressures takes on a log-Gaussian form that can be well described.
These results are compared to observations in
\S~\ref{sec:observations}, and the paper is summarized in
\S~\ref{sec:summary}.

\section{Models}

\label{sec:models}

The calculations presented here were done using the Riemann framework
for computational astrophysics, which is based on higher-order Godunov
schemes for MHD (Roe and Balsara 1996; Balsara 1998a,b; Balsara 2004),
and incorporates schemes for pressure positivity (Balsara \& Spicer
1999), and divergence-free magnetic fields (Balsara 2004). Balsara \&
Kim (2004) calibrated the method for the problem computed here.

We solve the ideal MHD equations including both radiative cooling and
uniform heating, but neither self-gravity nor external gravity.  Our
computational domain is a (200 pc)$^3$ periodic cube, large enough to
contain multiple SN remnants, but small enough to maintain some
contact with turbulence in real, stratified galactic disks. As we do
not include stratification, we must view our models as physics
experiments relevant to the ISM, rather than truly self-consistent
models.  We use grid resolutions of 64$^3$ and $128^3$ cells.  The
simulations are started with a uniform density of $2.3 \times
10^{-24}$ g cm$^{-3}$, threaded by a uniform magnetic field in the
$x$-direction with strength either 2~$\mu$G or 5.8 $\mu$G, to cover
the range of fields suggested by observations of the Milky Way disk
(Rand \& Kulkarni 1989, Beck 2001).  Behavior that is common to both
sets of models can be deduced to be fairly independent of the magnetic
field.

For the cooling, we use a tabulated version of the radiative cooling
curve shown in Figure~1 of MacDonald and Bailey (1981).  This curve
follows the equilibrium ionization cooling curve of Raymond, Cox \&
Smith (1976) for $10^6 \mbox{ K} < T < 10^8$~K; the nonequilibrium
curve for isochoric cooling derived by Shapiro and Moore (1976) from
$10^4 \mbox{ K} < T < 10^6$~K; and then is smoothly extrapolated to
zero at $10^2$~K.  It does not incorporate a sharp cutoff at $10^4$~K
due to the turnoff of Ly$\alpha$ cooling, nor does it include the
expected region of thermal instability below $10^4$~K.  We also
include a diffuse heating term to represent processes such as
photoelectric heating by starlight, which we set constant in both
space and time.  We set the heating level such that the initial
equilibrium temperature determined by heating and cooling balance is
3000~K.  Since the cooling time is usually shorter than the dynamical
time, we adopt implicit time integration for the cooling and heating
terms.

We set up SN explosions by adding 10$^{51}$ erg of thermal energy into
a sphere with radius 5 pc.  The SNe explode at random positions.
Unlike Korpi et al.\ (1999), we allow SNe to explode in regions of any
density to account for the dominant population of B stars (McCray \&
Kafatos 1987) that last far longer than their parent molecular clouds
(Fukui et al.\ 1999).  This allows SNe to explode within the shells of
pre-existing remnants. In our model they are not focussed into
associations, however.  Some properties of these models were described
in Balsara, Kim, \& Mac Low (2001).

\begin{figure}[htbp]  
 \centerline{\hbox{
\psfig{file=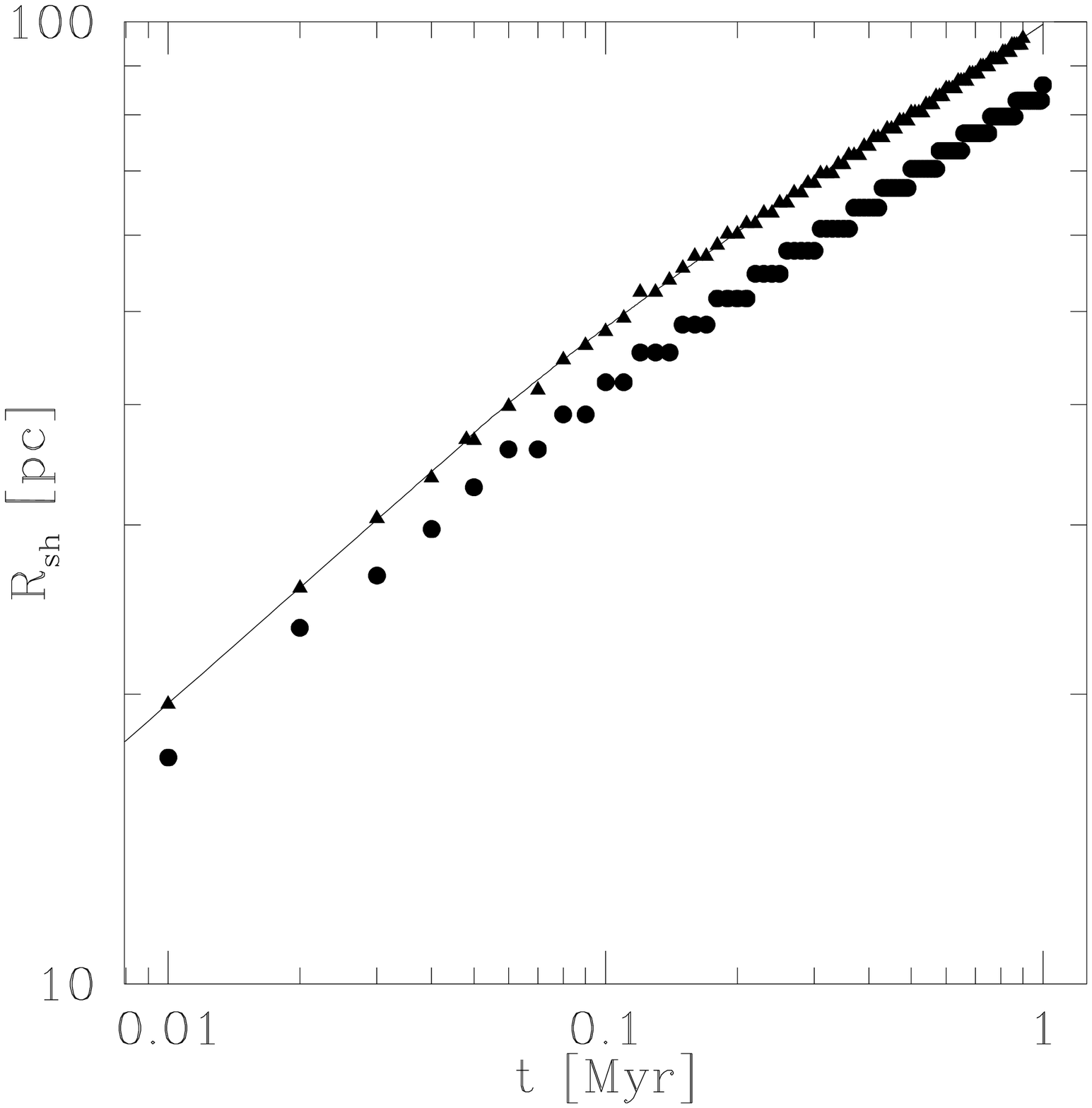,width=\textwidth}
}}
  \caption{Convergence study of shell radius for a single SN remnant
    in a cold, unmagnetized medium.  The analytic expression from
    Cioffi et al.\ (1988) ({\em line}) is compared to simulations in a
    200~pc box using grids of $64^3$ ({\em circles}), $128^3$ ({\em
      triangles}), and $256^3$ ({\em squares}) zones. Good agreement
    is obtained at $128^3$ resolution. Stair steps in the numerical
    curves come from only being able to measure the size of the shell
    to the nearest integral zone.}
  \label{fig:conv}
\end{figure}

To test the properties of the code, we set up an isolated SN explosion
in an unmagnetized medium with the properties of the test problem
described by Cioffi, McKee, \& Bertschinger (1988).  We ran this model
using the current code and cooling curve until the shell hit the edge
of our 200~pc box. Figure~\ref{fig:conv} shows tests with $64^3$,
$128^3$ and $256^3$ 
zones demonstrating that already with 3.13~pc resolution, an
isolated explosion agrees excellently with the analytic description of
Cioffi et al.\ (1988).

Table~\ref{tab:runs} gives the parameters of our runs, including SN
rate in terms of the Galactic rate, grid resolution $\Delta x$, and
initial uniform field strength $B_0$.  We take the Galactic SN rate,
scaled to the area of our grid, to be (1.2 Myr)$^{-1}$.  The final rms
magnetic field $<B>$, and the rms strength of the disordered field
$\langle \delta B \rangle$ are listed along with their ratio, which is
discussed in \S~\ref{subsec:therm}.  We also list rms Mach number and
variance of pressure in our runs for comparison with analytic
descriptions of turbulence in \S~\ref{subsec:pdfs}.

The evolution of the system is determined by the energy input from SN
explosions and diffuse heating and the energy lost by radiative
cooling.  We follow the simulations until the total energy of the
system, as well as the energy in the thermal, kinetic and magnetic
energies, has reached a quasi-stationary value for several million
years.  Similar runs with much weaker initial magnetic fields have
been used to study the amplification of magnetic fields in
interstellar turbulence by Balsara et al.\ (2004).  The fields in our
runs are amplified only modestly, as can be seen by comparing $B_0$ to
$\langle B \rangle$ in Table~\ref{tab:runs}, because we started our
runs with relatively strong initial fields.

\section{Analytic Theories}
\label{sec:an-th}

\subsection{Non-interacting Supernova Remnants}
\label{sub:snrs}
The two-phase theory of FGH assumed pressure equilibrium throughout
the ISM, with densities and temperatures fully regulated by heating
and cooling processes acting on timescales shorter than the dynamical
timescale of the gas.  The introduction of SN explosions by Cox \&
Smith (1974) and MO77 required relaxation of the assumption of pressure
equilibrium, at least inside of expanding SNRs.  In particular,
Jenkins et al.\ (1983) emphasized that MO77 implicitly makes
a prediction of the spectrum of pressure fluctuations expected from
the passage of SNRs expanding in a clumpy medium.

The theory of pressure fluctuations begins from the scaling in MO77,
Appendix B, of the pressure $P$ in the hot ionized medium, which is
given as a function of the probability $Q(R)$ of a point being within
a SNR with radius at most $R$,
\begin{equation} \label{qeqn}
P = P_{c} \left\{\begin{array}{ll} 
       (Q/Q_c)^{-9/14}    & \mbox{ for $Q \leq Q_c$} \\
       0.5 (Q/Q_c)^{-0.9} & \mbox{ for $Q > Q_c$},    \end{array} \right.
\end{equation}
where $Q_c$ is the probability of a point being within a SN remnant
large enough for the swept-up shell of intercloud medium to have
cooled, and $P_c$ is the corresponding pressure of such a remnant.
Note that the immediate loss of pressure after shell cooling produces
an absence of pressures with $0.5 P_c < P < P_c$ in this model.
For typical values in the Milky Way, including a SN
rate $S = 10^{-13}$~pc$^{-3}$~yr$^{-1}$, MO77 find by balancing a number
of observational considerations that likely values for the
parameters in equation~(\ref{qeqn}) are $Q_c = 1/2$, and $P_{c}/k =
10^{3.67}$~cm$^{-3}$~K (see MO77, eq.~9). 

Equation~(\ref{qeqn}) can then be inverted 
to give the probability $Q(P)$ of a point being within a SN
remnant with pressure at least $P$, as written in Jenkins et al.\
(1983).
The probability $Q(P)$ represents a cumulative distribution.  The
corresponding differential PDF is given
by $-dQ(P)$.  In practice, we compute PDFs by constructing a histogram
of pressure values, with bins of finite size $\Delta P$, so the PDF
predicted by MO77 is 
\begin{equation} \label{mo77}
-(\partial Q/\partial P)\Delta P = \left\{\begin{array}{ll}
    \frac{14Q_c}{9P_c}\left(\frac{P}{P_c}\right)^{-23/9}\Delta P 
         & \mbox{for $P>P_c$},\\
    \frac{20Q_c}{9P_c}\left(\frac{P}{P_c}\right)^{-19/9}\Delta P 
         & \mbox{for $P \leq P_c$}.  \end{array} \right.
\end{equation}
This PDF would diverge towards low pressures if it were not limited by
the consideration that the pressure in an isolated SN remnant will not
fall below the ambient pressure $P_0$, so there must be a lower
cutoff at that value.  (MO77 give $P_0/k = 10^{3.10}$~cm$^{-3}$~K for
the same parameters given above.)  This model therefore predicts no
pressures below the ambient value $P_0$.

\subsection{Turbulence}
\label{sub:turb}

An alternative approach to predicting the distribution of interstellar
pressure fluctuations can be derived from work on properties of highly
compressible turbulence by PNJ97 and PV98\footnote{PV98 suffers from a
  number of typographical errors produced by a last-minute switch of
  notation (Passot, priv. comm.) to distinguish their scaling
  parameter $M$ from their rms Mach number $\tilde{M}$, which we
  denote as $M_{\rm rms}$.  We here enumerate the errors that we are
  aware of: (1) In the paragraph above their eq.~(17), $\tilde{M}$
  should be used on every occasion. (2) In the text immediately below
  their Figure 3, $\sigma_s = \tilde{M}$, not $M$. (Note however, that
  their Figure 3 is indeed labelled with $M$, not $\tilde{M}$.) (3)
  There is an extra $M$ in the first expression of the middle line of
  their equation 18, which should be simply $u_{\rm rms}/c(s)$. (4)
  The first term in the bracketed exponential in their equation (20)
  should contain an additional factor of $u_{\rm rms}^2$ in the
  denominator.}  inspired by the theory of V\'azquez-Semadeni (1994).
These papers considered the PDF of density ${\cal P(\rho)}$ in a
turbulent, isothermal gas with pressure $P = \rho c_s^2$. (PV98 and
Nordlund \& Padoan [1999] have also considered polytropic equations of
state, $P=K\rho^\gamma$ where $\gamma$ is the polytropic index, but
Nordlund \& Padoan [1999] note that the distribution shifts by less
than a factor of two even for polytropic index $\gamma = 0.3$, so we
only consider the isothermal case here.) By assuming that successive
passages of shocks and rarefaction waves acting as a random
multiplicative process build up density fluctuations
(V\'azquez-Semadeni 1994), they showed that the density distribution
is a log-normal
\begin{equation}
{\cal P}(s)ds = \frac{1}{\sigma\sqrt{2\pi}} \exp \left[ -
\frac{(s-s_0)^2}{2\sigma^2} \right] ds,
\end{equation}
where the variable $s = \ln \rho/\rho_0$, and $\rho_0$ is the mean
density of the region. 

The actual value of the variance of the logs of the densities $\sigma$
was extracted from numerical models assuming an isothermal equation of
state by both groups.  Using 3D models that have not been completely
described in the literature, PNJ97 found
\begin{equation}\sigma^2 = \ln(1 + M_{\rm rms}^2/4),
\label{eq:pnj97}
\end{equation}
while using very high resolution 1D models,
PV98 found 
\begin{equation}\sigma = M_{\rm rms},
\label{eq:pv98}
\end{equation}
where the scaled root mean square (rms) Mach number $M_{\rm rms} =
v_{\rm rms} / c_s$, the ratio of the rms velocity to the isothermal sound
speed.

From this formalism, we can derive the equivalent PDF 
in pressure ${\cal P}(P)$ by using the isothermal equation of state $P =
\rho c_s^2$. For convenience, we define $x = \ln P$, so that
\begin{equation}
s = x - \ln c_s^2,
\end{equation}
Then the isothermal pressure PDF is
\begin{equation}\label{eq:pdfiso}
{\cal P}(x)dx = \frac{1}{(2\pi\sigma^2)^{1/2}} \exp\left\{ -
\frac{ (x-x_0)^2}{2\sigma^2} \right\} dx,
\end{equation}
where $x_0 = \ln P(\rho_0)$.
The dispersion of the normalized natural logs of the pressures is thus
predicted to be the same as that for the pressures, as given in
equations~\ref{eq:pnj97} and~\ref{eq:pv98}. 
We compare our numerical results to these predictions in
\S~\ref{subsec:pdfs}.

\section{Numerical Results}
\label{sec:numerics}

\subsection{Morphology}

\begin{figure} 
\centerline{\hbox{
\psfig{file=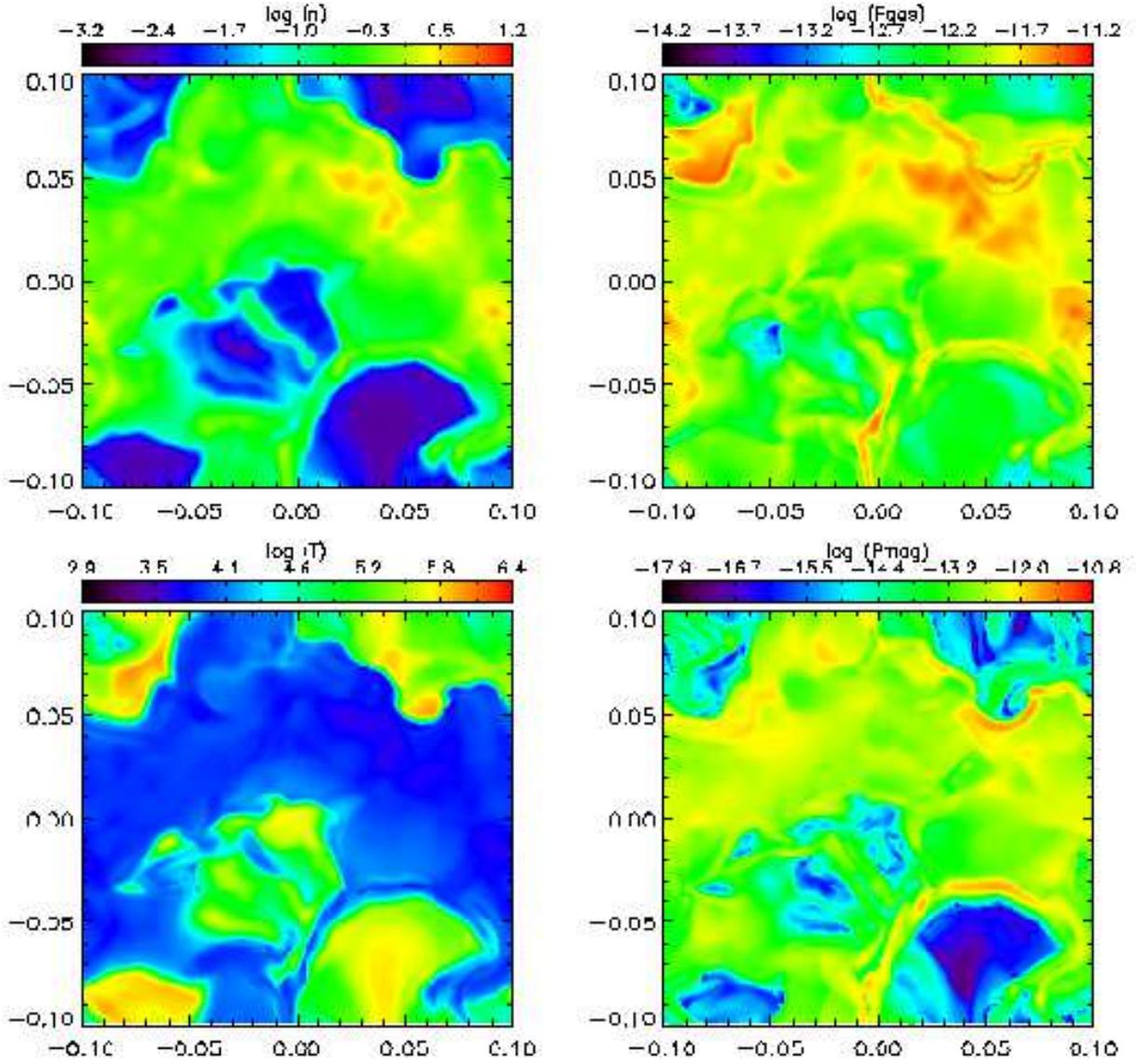,width=\textwidth}
}}  
\caption{\label{fig:mhd-cut} Two-dimensional slices through the
three-dimensional model M8, parallel to the magnetic field at a
time of 25~Myr, showing density (upper left), thermal pressure
(upper right), temperature (lower left), and magnetic pressure (lower
right). Color bars indicate the scale of each quantity, in units of
cm$^{-3}$ for density $n$, K for temperature $T$, and dyne~cm$^{-2}$
for both pressures.  The axes are labeled in units of kpc.
}
\end{figure}

We now examine the pressure distribution in our numerical simulations
of a SN-driven, magnetized, interstellar medium.  In
Figure~\ref{fig:mhd-cut} we show density, pressure, and temperature,
as well as magnetic pressure, on cuts through model M2 parallel to the
magnetic field. (The perpendicular direction appears identical,
because the flows are strongly super-Alfv\'enic, with rms Alfv\'en
number exceeding four.) Examination of the pressure images immediately
shows a broad variation in pressures among different regions in all
the models, including in regions not closely associated with young
SNRs.  Regions with pressures markedly lower than ambient are
apparent.

High temperature regions lie inside young SNRs, while low temperature
regions have no uniform density and temperature correlation.  The
highest density regions have sizes of dozens of parsecs, and average
densities approaching 100~cm$^{-3}$, typical of giant molecular
clouds.  This is consistent with the suggestion by Ballesteros-Paredes
et al.\ (1999a) and Ballesteros-Paredes, V\'azquez-Semadeni \& Scalo
(1999) that at least smaller molecular clouds are formed and destroyed
by the action of the interstellar turbulence.  Our models do not,
however include self-gravity 
as do Li, Mac Low, \& Klessen (2005) or
Slyz et al.\ (2005), though neither includes magnetic fields (for
recent reviews see Elmegreen 2002;  
Mac Low \& Klessen 2004), isobaric thermal instabilities (Hennebelle
\& P\'erault 1999, 2000; Burkert \& Lin 2000; V\'azquez-Semadeni,
Gazol, \& Scalo 2000; and Gazol et al.\ 2001), or thermal conduction
(Koyama \& Inutsuka 2004).

Even at SN rates four times those characteristic of the Milky Way, the
hot medium in our models does not surround discrete clumps of cold and
warm gas.  Rather, discrete regions of hot gas are formed,
occasionally intersect, and then seem to be dynamically mixed back
into the warm gas that fills a substantial fraction of the space.
This large-scale turbulent mixing appears to enhance the cooling rate,
while sheets and filaments confined by nearby SNRs seem more effective
at slowing down the expansion of SNRs than 
the isolated, discrete clumps of the same equivalent density
suggested by MO77.  
These results are consistent with the
results of Korpi et al.\ (1999), Avillez (2000),  Avillez \&
Breitschwerdt (2004),
and Slyz et al.\ (2005).

\subsection{Thermodynamic Relations}
\label{subsec:therm}

The first theories of the multi-phase ISM, such as FGH, postulated an
isobaric medium.  Since then, multi-phase models have commonly been
interpreted as being isobaric, although MO77 and Wolfire et al.\ (1995)
actually assume only local pressure equilibrium, not global, and MO77
did consider the distribution of pressures, as described above in
\S~\ref{sub:snrs}.  In multi-phase models, the heating and cooling
rates of the gas have different dependences on the temperature and
density, so that the balance between heating and cooling determines
allowed temperatures and densities for any particular pressure.  This
balance can be shown graphically in a phase diagram, showing, for
example, the allowed densities for any pressure (FGH; for a modern
example, see Fig.~3({\em a}) of Wolfire et al.\ 1995).  

\begin{figure} 
\centerline{\hbox{
\psfig{file=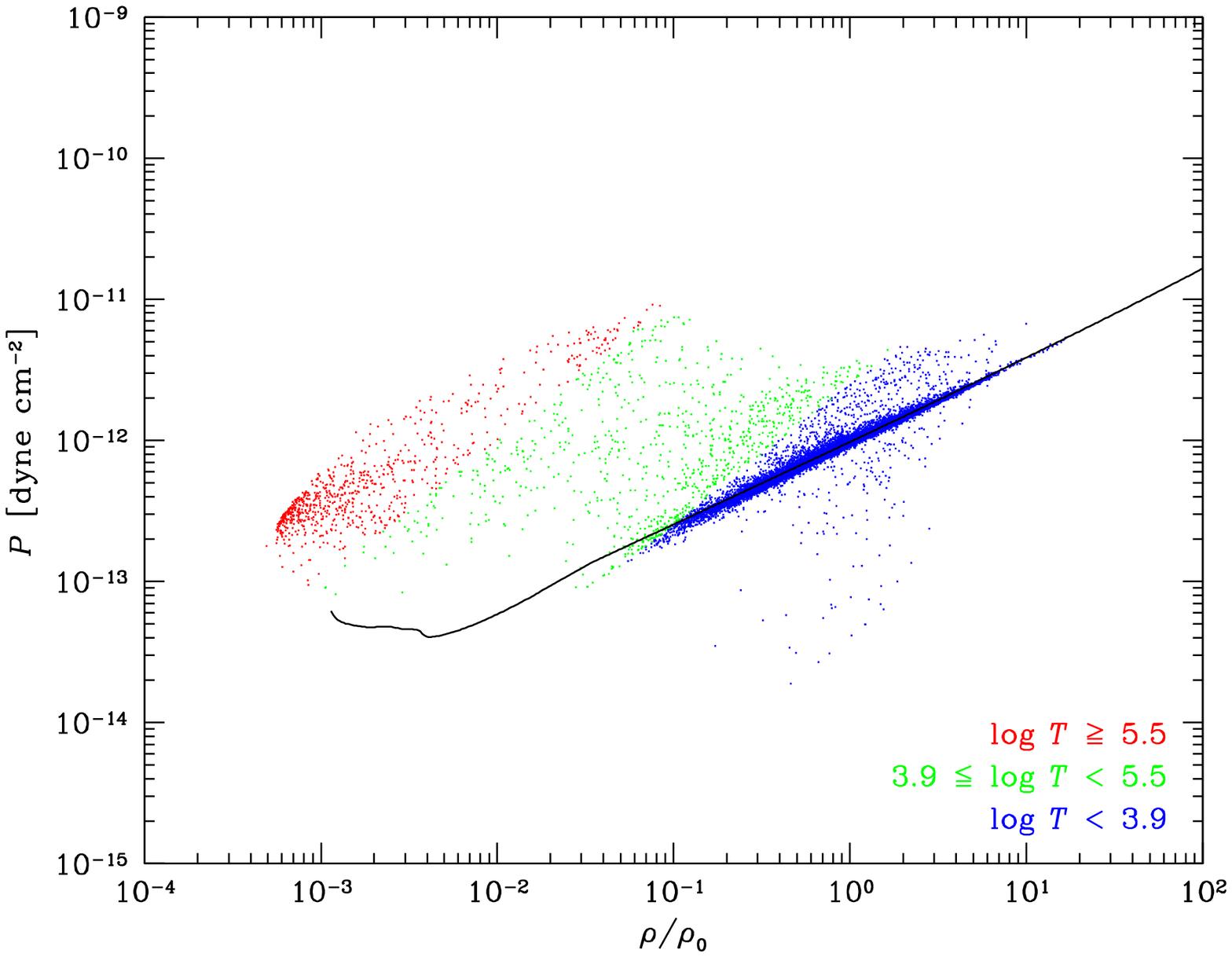,width=0.5\textwidth}
\psfig{file=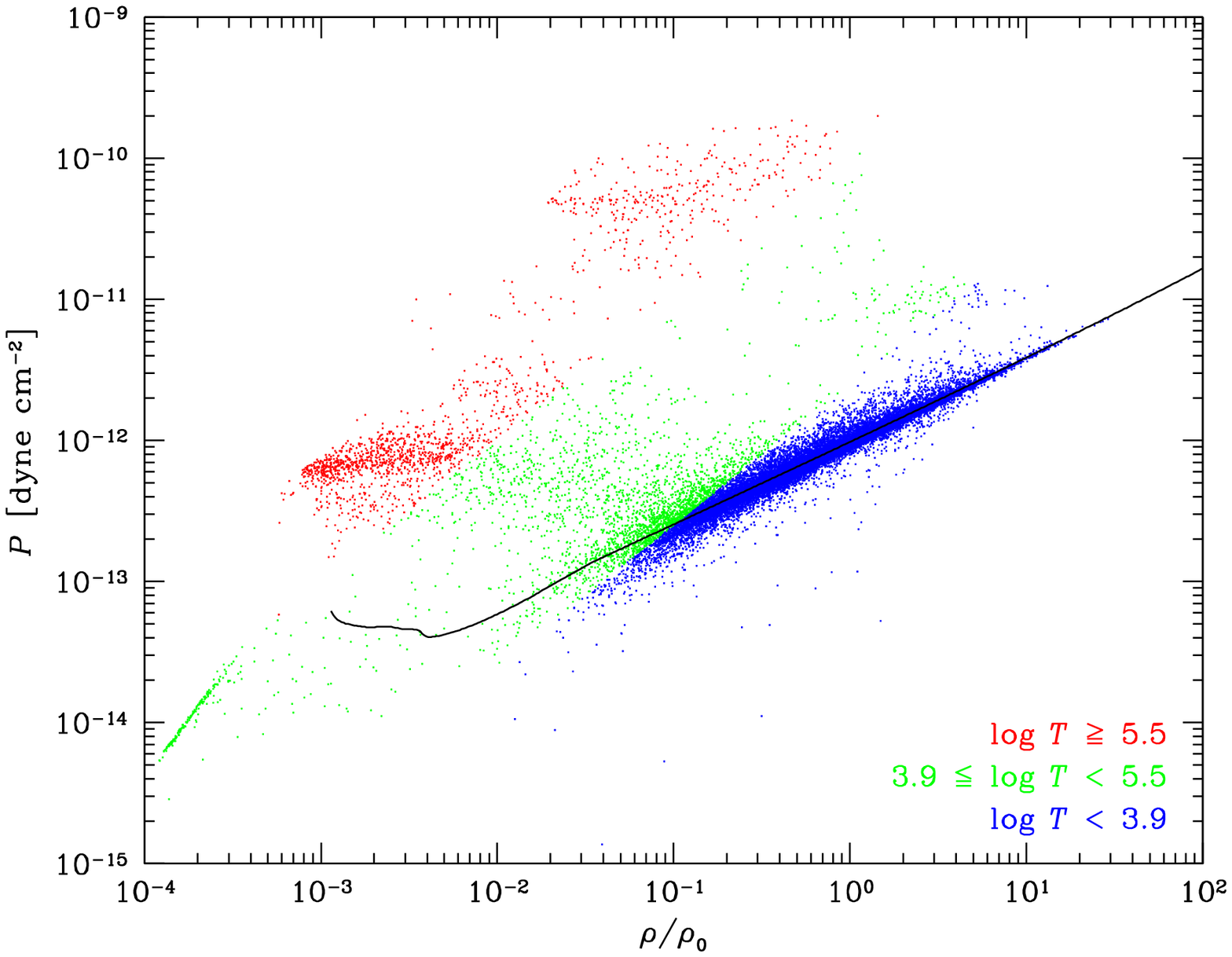,width=0.5\textwidth}
}} \centerline{\hbox{
\psfig{file=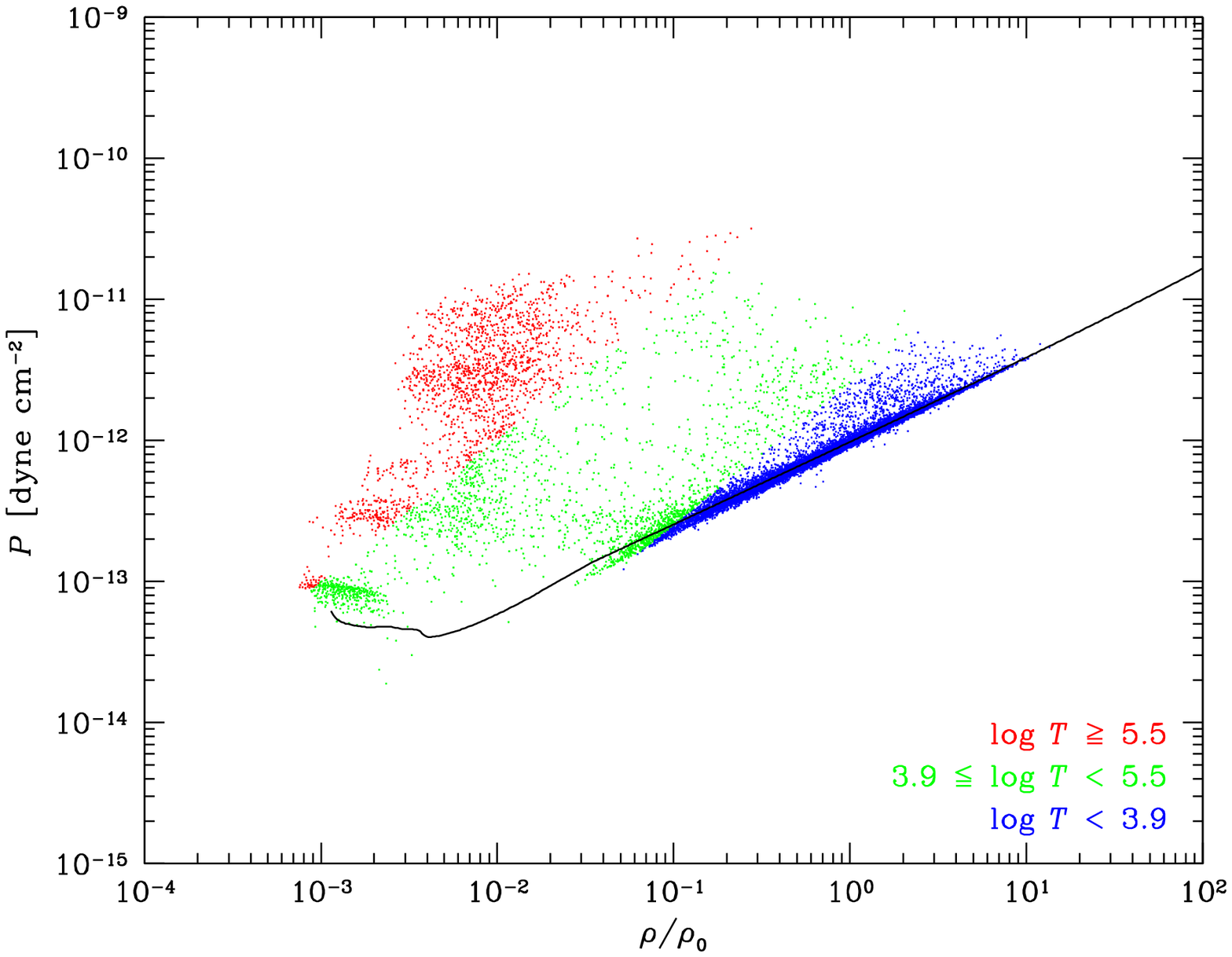,width=0.5\textwidth}
\psfig{file=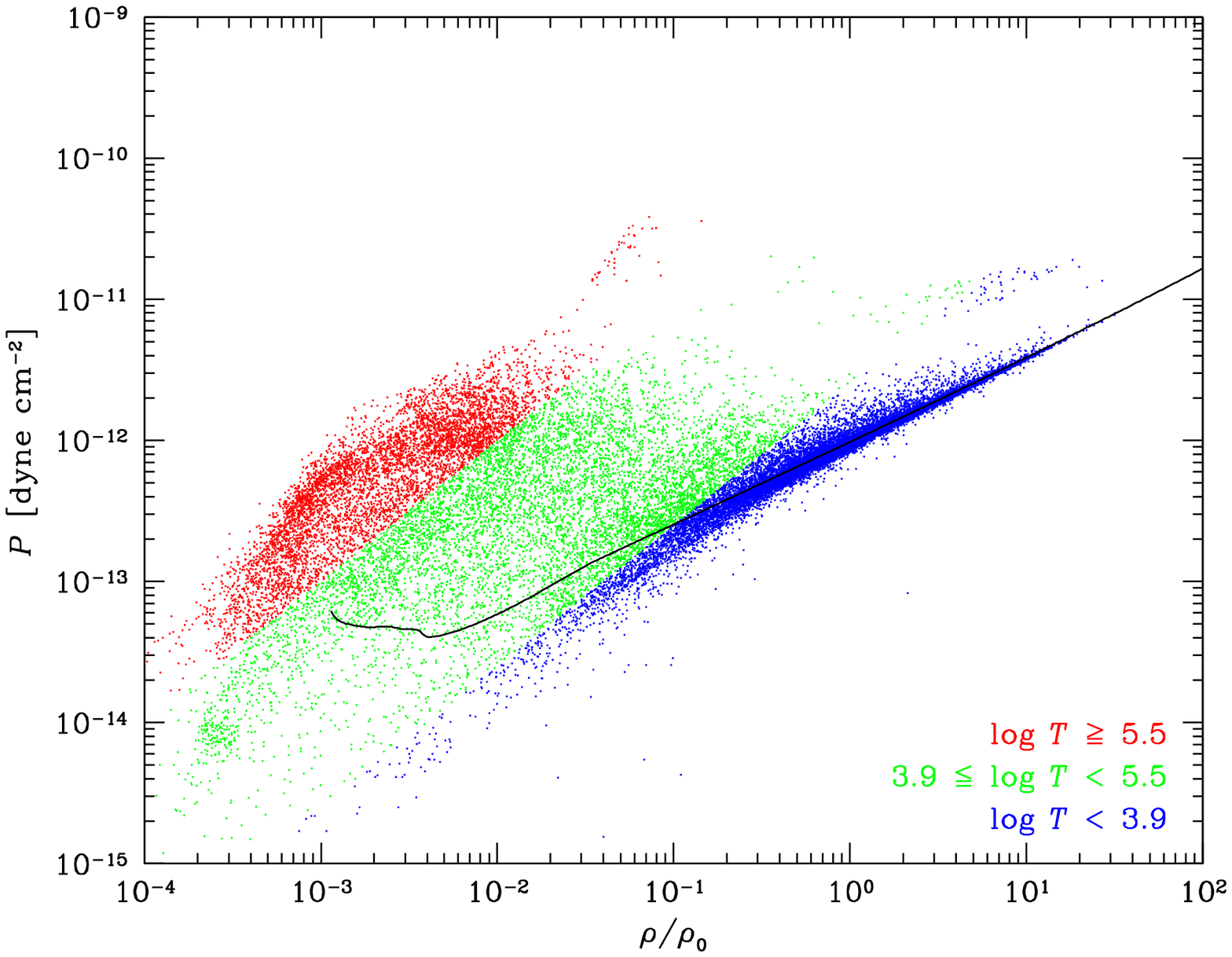,width=0.5\textwidth}
}}
\caption{\label{fig:mhd-dscat}
  Scatter plot of pressure vs.\ density at $t=25$ Myr in models {\em
    (a)} M2, with low SN rate and high field, {\em (b)} M4, with high
  SN rate and field, {\em (c)} M6, with low SN rate and field, and
  {\em (d)} M8 with high SN rate and low field.  We show a subset of
  $32^3$ points sampled at intervals of eight points in each
  direction, at 25 Myr for models M4 and M8, and 100 Myr for models M2
  and M6.  Cool gas with $\log T < 3.9$ is shown in blue, warm gas
  with $3.9 < \log T < 5.5$ in green, and hot gas with $\log T > 5.5$
  in red. Temperature is given in K.  The thermal equilibrium curve
  for the cooling and heating functions in this simulation is overlaid
  as a black line.  Note that a wide variation of pressures is seen,
  and for each pressure, different densities occur.  }
\end{figure}

In Figure~\ref{fig:mhd-dscat}, the thermal-equilibrium curve for the
heating and cooling mechanisms included in the models is shown as
a black line. Only
a single phase is predicted at high densities as our cooling curve did
not include the physically-expected unstable region at temperatures of
order $10^3$~K (Wolfire et al.\ 1995).  Thus, if our model produced an
isobaric medium, it would be expected to have a single low-temperature
phase at uniform density given by the point at which the
thermal-equilibrium curve crosses that pressure level.  (Effectively,
our cooling curve allows the hotter two of the three phases proposed by MO77.)

Figure~\ref{fig:mhd-dscat} show the actual density and pressure of
individual zones in the model. Many zones at temperatures in the range
of $10^3$--$10^4$~K do lie on the thermal equilibrium curve, but
scattered up and down it at many different pressures and densities,
with no well-defined phase structure.  Furthermore, a substantial
fraction of the gas has not had time to reach thermal equilibrium at
all after dynamical compression.  It appears that pressures are
determined dynamically, and the gas then tries to adjust its density
and temperature to reach thermal equilibrium at that pressure.  Most
gas will land on the thermal equilibrium curve when dynamical times
are long compared to heating and cooling times.  This still leads to
all points within the range of pressures available along the thermal
equilibrium line being occupied, rather than the appearance of
discrete phases.  Unstable regions along the thermal equilibrium curve
(Gazol et al.\ 2001) and off it will also be populated, as observed by
Heiles (2001) for colder gas, but not as densely, as gas indeed
attempts to heat or cool to a stable thermal equilibrium at its
current pressure.

Comparison of the models shown in Figure~\ref{fig:mhd-dscat}{\em
  (a)--(d)} shows that the amount of gas that has not yet reached
thermal equilibrium depends strongly on the SN rate, but that the
basic conclusion that gas ends up in thermal equilibrium at a wide
range of pressures does not. Magnetic fields play a secondary role,
with higher field strength increasing the total pressure, but somewhat
reducing the range of thermal pressures occupied.  A stronger magnetic
field does produce regions of low thermal pressure balanced by
magnetic pressure that can be seen on and below the thermal
equilibrium curve, as we further discuss below.

The range of pressures observed in our simulations is
broader than the region shown by Wolfire et al.\ (1995) to be subject
to thermal instability.  Even models that included a proper cooling
curve would produce some gas at pressures incapable of supporting a
classical multi-phase structure.  The mixture of different pressures
will, however, produce gas at both high and low densities, as well as
a smaller fraction of gas at intermediate densities that has not yet
reached thermal equilibrium.

Under what conditions are the dynamical times indeed long compared
to heating and cooling times?  We can compare
rough analytical estimates of each.  The dynamical time is
\begin{equation}
t_{\rm dyn} = L/v_{\rm rms},
\end{equation}
where $L$ is a characteristic length scale, while the cooling time 
\begin{equation}
t_{\rm cool} = E/\dot{E} = kT/ n \Lambda,
\end{equation}
where $E$ is the thermal energy, and $\dot{E} = n^2 \Lambda(T)$ is the
cooling rate as a function of temperature $T$. 

In model M2, for
example, the rms velocity volume-averaged over gas of all temperatures
is $v_{\rm rms} = 55 \mbox{ km s}^{-1}$.  (This is the model with the
smallest value of $v_{\rm rms}$.)  If we take typical dynamical length
scales of $L\sim 10$~pc, then $t_{\rm dyn} \simeq 0.2$~Myr.  We
tabulate $t_{\rm cool}$ from our cooling curve in
Table~\ref{tab:cool}, normalized to a density of $n=1$~cm$^{-3}$. (We
note that model cooling times in gas at temperatures of $10^3\mbox{ K}
\lesssim T \lesssim 10^4$~K are substantially less than physical
values, as the MacDonald \& Bailey cooling curve does not drop
abruptly at $10^4$~K when Ly$\alpha$ cooling shuts off.  However, the
consequence of this is merely that our model overestimates the amount
of gas that has reached thermal equilibrium: the scatter plot shown in
Figure~\ref{fig:mhd-dscat} should be even more uniformly filled.) At
all temperatures $10^3$~K$ < T < 10^7$~K, we find that $t_{\rm cool}
\ll t_{\rm dyn}$, especially at densities $n>1$~cm$^{-3}$.  This
agrees with our description of gas dynamics setting the local pressure
and thermal equilibrium only then determining the density and
temperature.

The separation between dynamical and thermal timescales also sheds
light on the study by V\'azquez-Semadeni et al.\ (2000) on the effects
of turbulence on thermal instability.  That study suggested that
turbulence erases the effects of thermal instability on the
interstellar medium.  Our models show that
turbulence probably didn't erase the thermal instability entirely, but
that the instability only acted locally, under the conditions set for
it by the larger-scale turbulent flow.  Also, thermal instability
becomes less important in determining the overall distribution of
pressures and temperatures when much of the gas has not even reached
thermal equilibrium.  More gas will lie in thermally stable regions
than thermally unstable ones, but the wide range of available
pressures and the lack of complete thermal equilibrium in many regions
still results in a wide range of properties, despite the nominal
action of the thermal instability.

\begin{figure} 
\centerline{\hbox{
\psfig{file=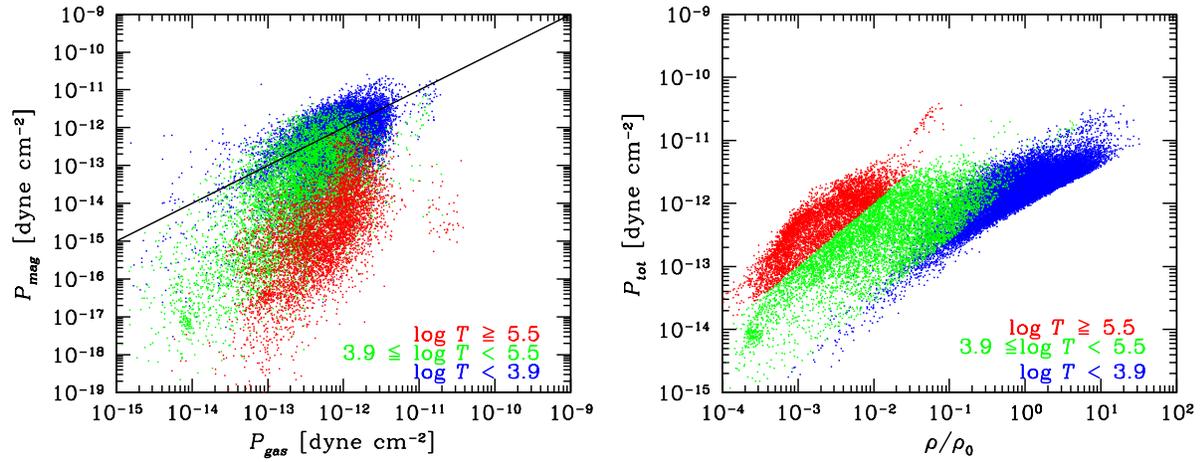,width=\textwidth}
}}
\caption{\label{fig:mhd-mscat} Scatter plot of (a) magnetic vs.\ thermal
  pressure and (b) total pressure vs.\ density at $t=25$~Myr in model
  M8.  We again show a subset of $32^3$ points.  In ({\em a}), many of
  the blue cold gas points have been covered by green warm and red hot
  gas points. The solid line shows equipartition between magnetic and
  thermal pressure.  Much of the gas is far from equipartition in all
  temperature ranges.  
}
\end{figure}

The lower boundary of the heavily occupied region in the
pressure-density plane appears to be determined by the polytropic
behavior of the gas near this cutoff.  Fitting to its slope yields
values of the polytropic index $\gamma \sim$~0.6--0.7.  This can be
understood as due to a cooling curve increasing as a power law
(roughly $T^{2.9}$ can be fit) balanced by heating.
Points lying below and to the right of this cutoff line at low
pressures and high densities appear to be magnetically supported,
dropping to very low thermal pressures at intermediate densities.
This gas does not have low total pressure, only low thermal pressure,
as shown in Figure~\ref{fig:mhd-mscat}({\em b}). 

In Figure~\ref{fig:mhd-mscat}{\em (a)}, the relative strength of
thermal and magnetic pressure is shown at one time for a magnetized
simulation.  Equipartition of thermal and magnetic pressure (plasma
$\beta = P_{\rm gas} / P_{\rm mag} = 1$) falls along the solid line.
Gas at all temperatures often falls far from the equipartition line.
In this model, both cold and warm gas have values of $\beta$ as low as
0.1, while at the high end, cold gas reaches $beta \sim 100$, and warm
and hot gas have regions with $\beta$ as high as 10$^4$.  Magnetically
supported regions occur in both cold and warm gas, as do thermally
dominated regions.  Hot gas can be seen, on the other hand, to be
dominated by thermal pressure, with low magnetic pressures.  In
Figure~\ref{fig:mhd-mscat}{\em (b)}, the total pressure is shown as a
function of density for the same zones. Using total pressure, a clear
cutoff at high densities and low pressures is now plain to see,
showing that the previous scatter of points beyond it was due to small
magnetically supported regions.

The wide range of magnetic pressures shown in
Figure~\ref{fig:mhd-mscat}{\em a} calls into question the assumption
of equipartition between magnetic and cosmic ray energy usually used
to derive magnetic field strengths from radio synchrotron emission
(Burn 1966, Sokoloff et al.\ 1998, Beck 2001). Large fluctuations in
the magnetic field strength lead to overestimates of the strength of
the regular magnetic field for given strength of polarized synchrotron
emission.  Beck (2001) estimates that fluctuations $\langle \delta B
\rangle / \langle B\rangle \simeq 1$ lead to as much as a 40\%
overestimate in field strength or a factor of two overestimate in
field energy.  To measure the size of the fluctuations in our models,
we take $\langle B \rangle = B_{\rm rms}$, and the fluctuating field
$\langle \delta B \rangle$ to be the rms value of the field with the
initial mean field $B_0$ subtracted off before taking the rms. The
resulting values at the end of each of our runs, along with the ratio
$\langle \delta B \rangle / \langle B\rangle$ are reported in
Table~\ref{tab:runs}.  At lower field strengths, the fluctuations
indeed approach order unity, and even with the high field strengths
suggested by the synchrotron emission, $\langle \delta B \rangle /
\langle B\rangle > 0.5$.  These results suggest that synchrotron
emission observation may yield moderate to severe overestimates of the
total field strength.

\begin{figure}  
\centerline{\hbox{
\psfig{file=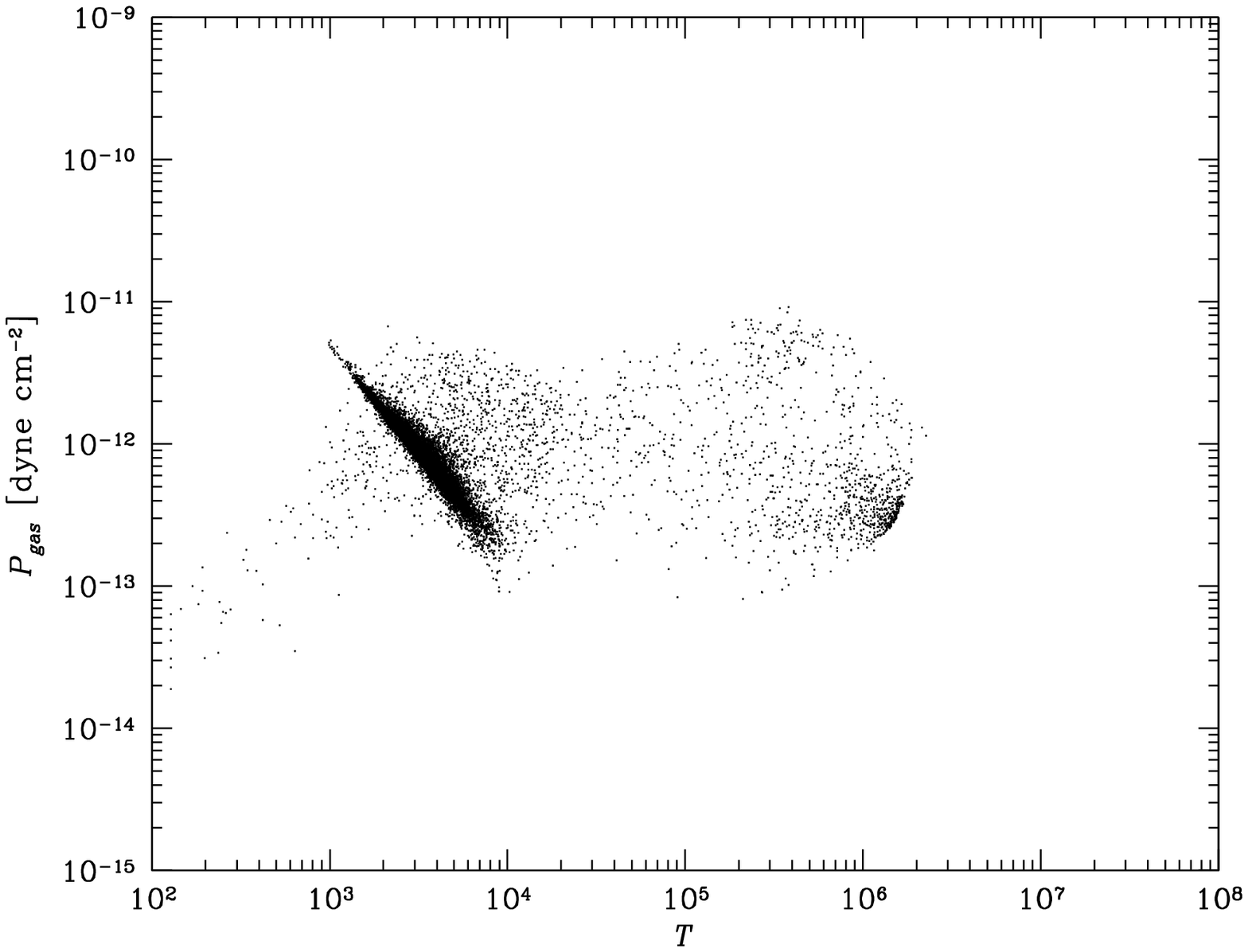,width=0.5\textwidth}
\psfig{file=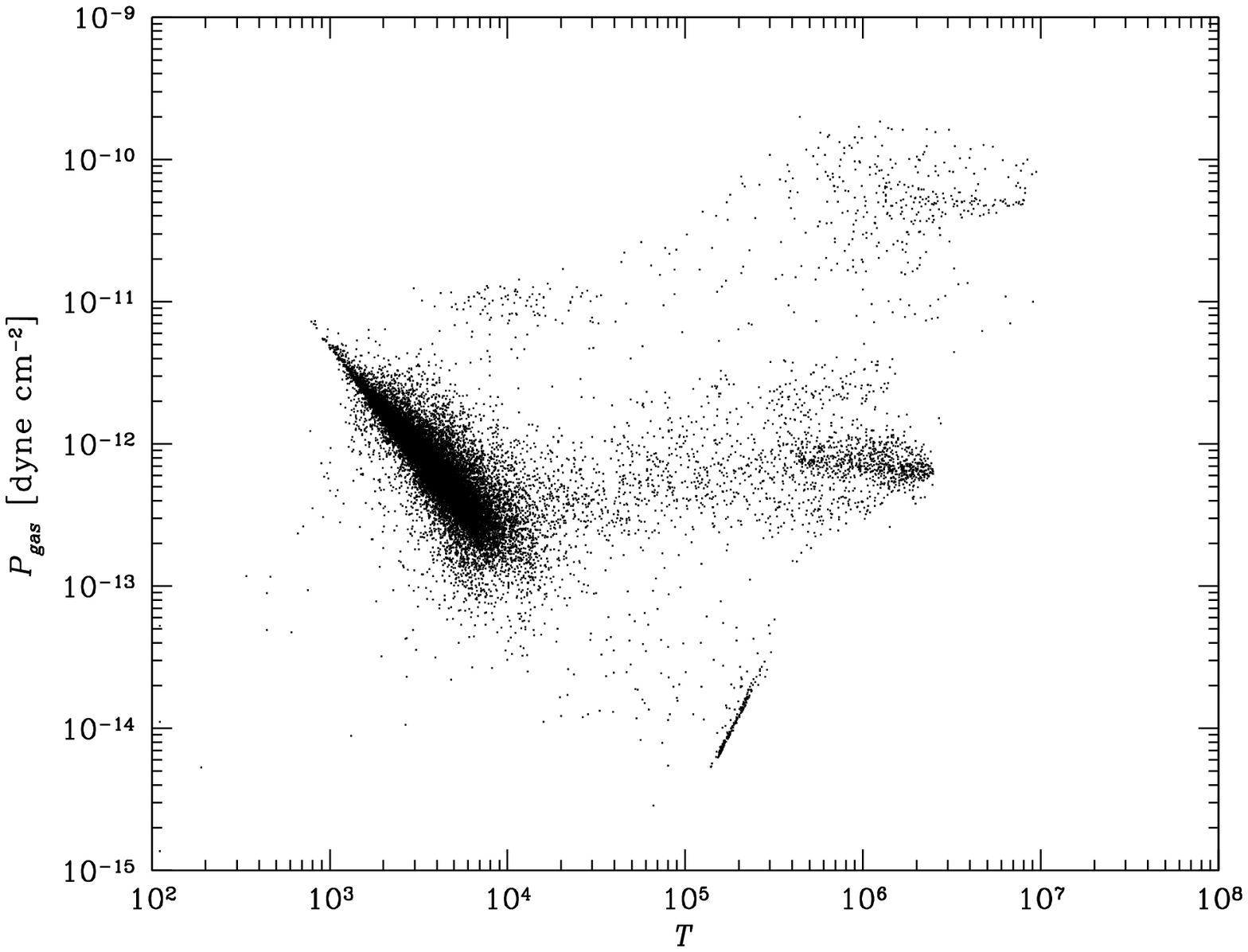,width=0.5\textwidth}
}} \centerline{\hbox{
\psfig{file=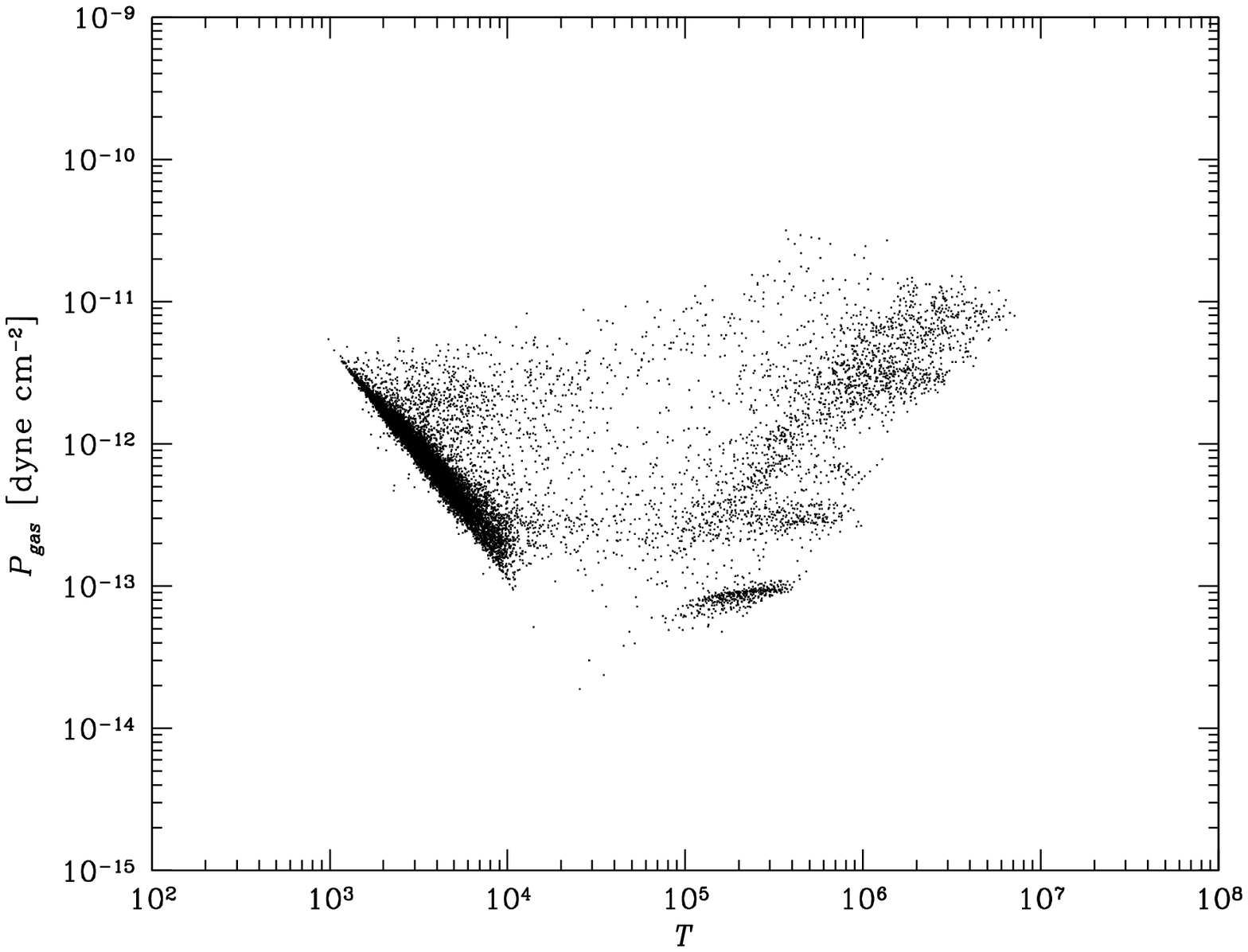,width=0.5\textwidth}
\psfig{file=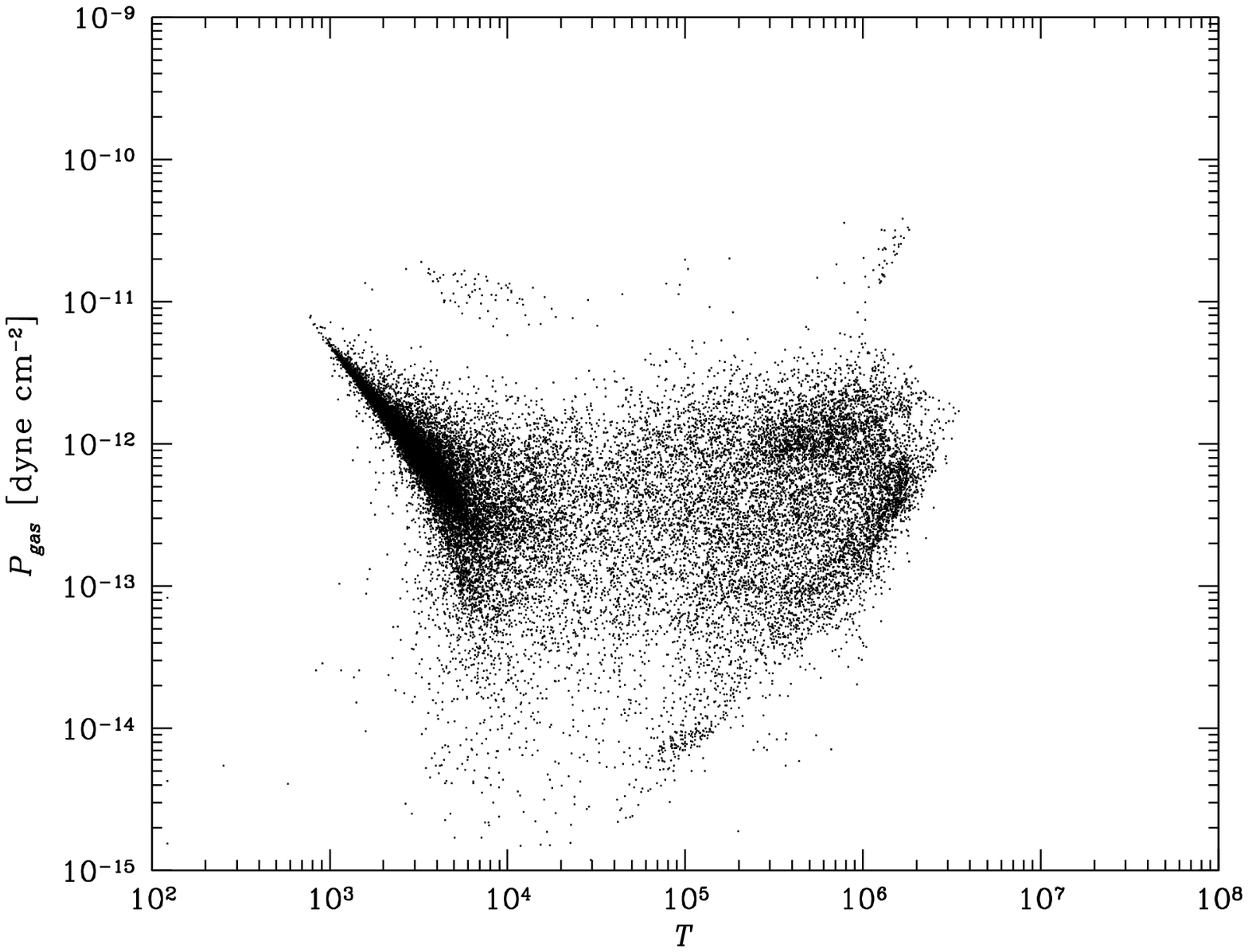,width=0.5\textwidth}
}}
\caption{\label{fig:mhd-tscat}
  Scatter plot of pressure vs.\ temperature at $t=25$ Myr in models
  {\em (a)} M2, with low SN rate and high field, {\em (b)} M4, with
  high SN rate and field, {\em (c)} M6, with low SN rate and field,
  and {\em (d)} M8 with high SN rate and low field.  We again show a
  subset of $32^3$ points, and give temperature in K.
}
\end{figure}

Finally, let us directly consider the distribution of pressures in gas
at different temperatures.  In Figure~\ref{fig:mhd-tscat} we show
scatter plots of thermal pressure against temperature.  We note the
concentration of points at temperatures below $10^4$~K, which once
again reflects gas piling up in thermal equilibrium at a wide variety
of pressures.  The tilt to the left in Figure~\ref{fig:mhd-tscat}
occurs from the effective adiabatic index $\gamma < 1$, with higher
pressure gas typically having lower temperatures.  Individual SNRs
with roughly constant pressures are visible as stripes in the higher
temperature region of the plot.  As they cool, they also expand to
lower pressures. Higher SN rate produces a broader distribution of
temperatures in the low field case, where SN bubbles expand to fill
more of the total volume.

\subsection{Model Probability Distribution Functions}
\label{subsec:pdfs}

\begin{figure}  
\centerline{\hbox{
\psfig{file=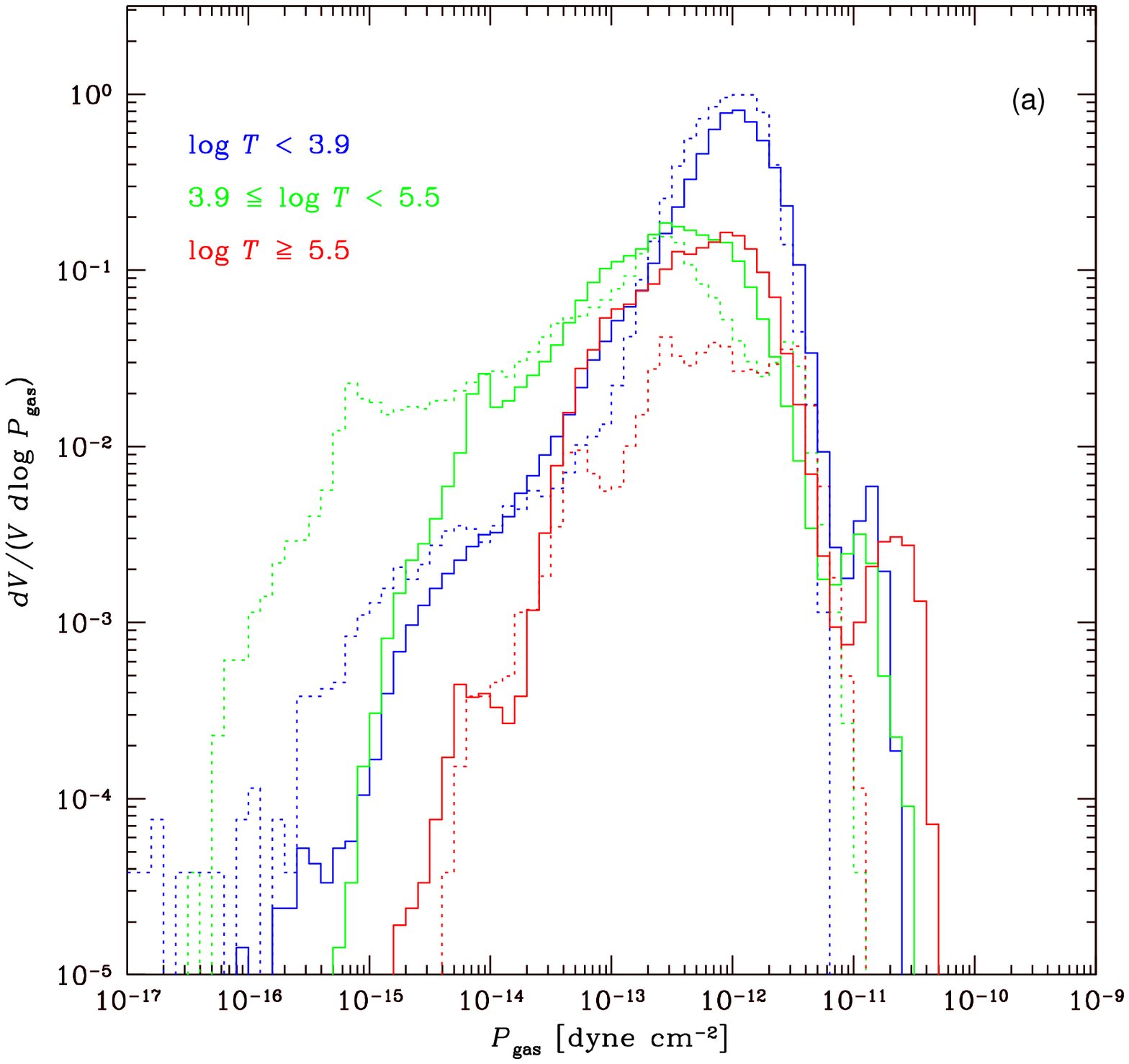,width=0.5\textwidth}
\psfig{file=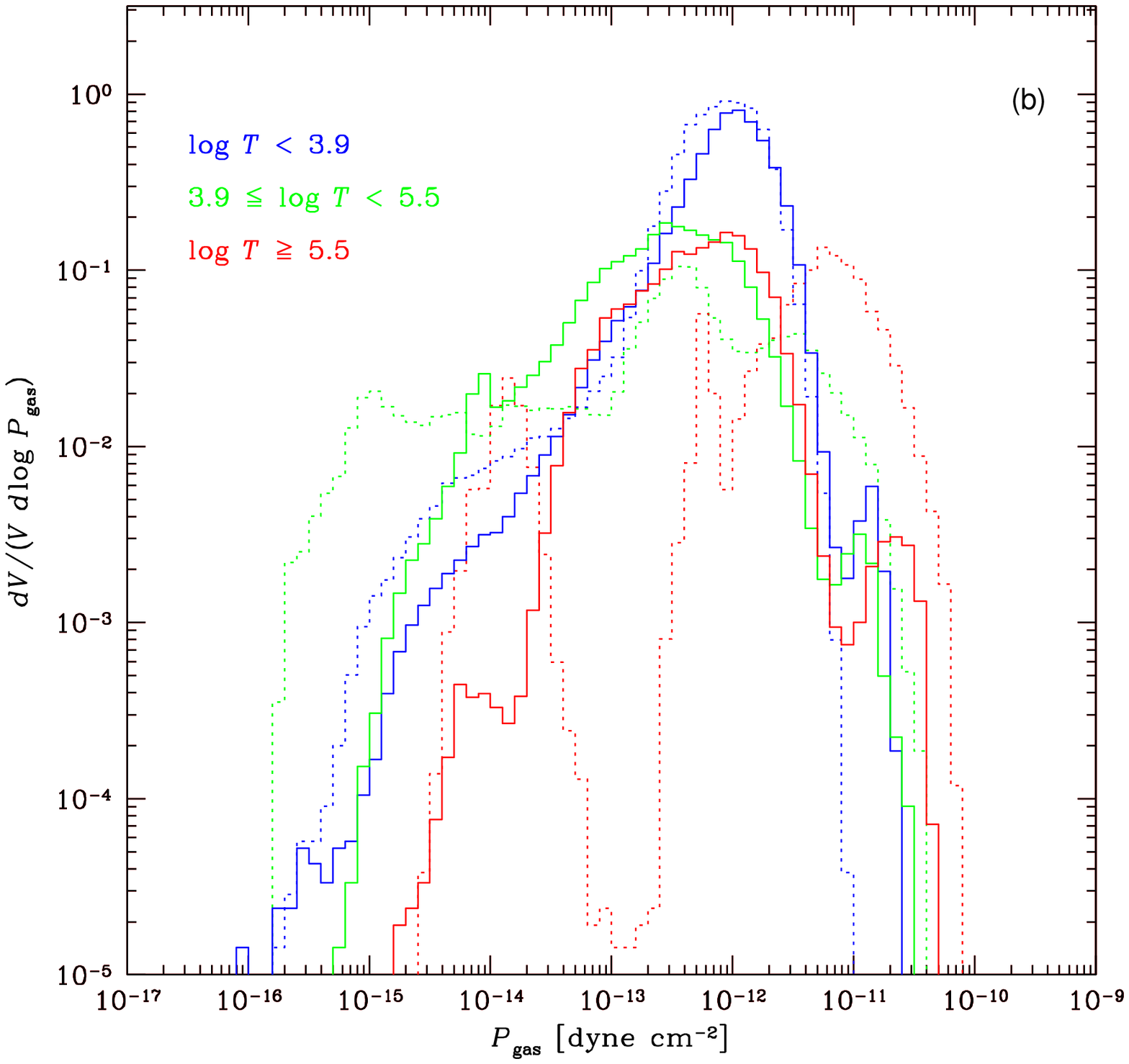,width=0.5\textwidth}
}}
\centerline{
\hbox{
\psfig{file=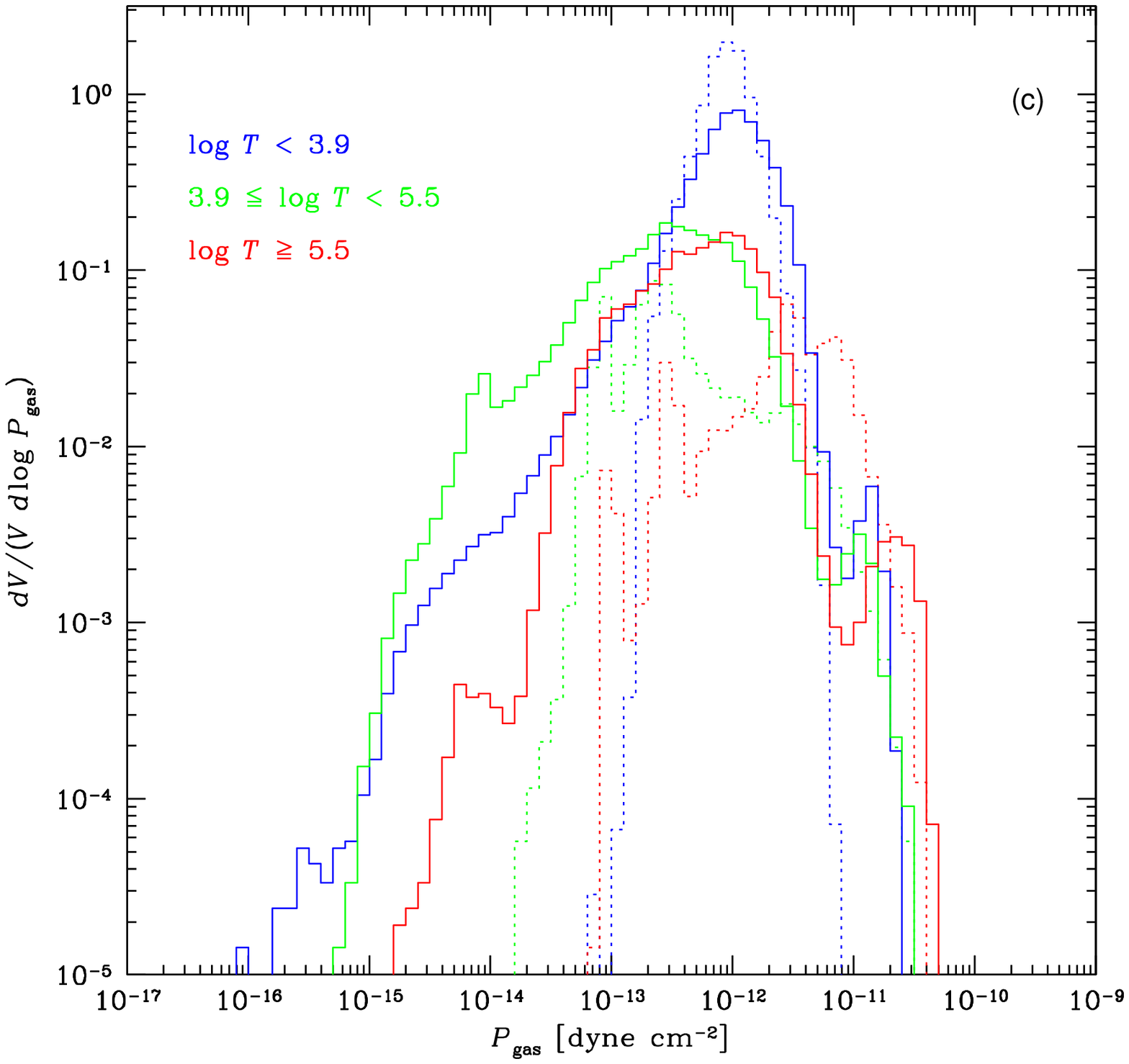,width=0.5\textwidth}
\psfig{file=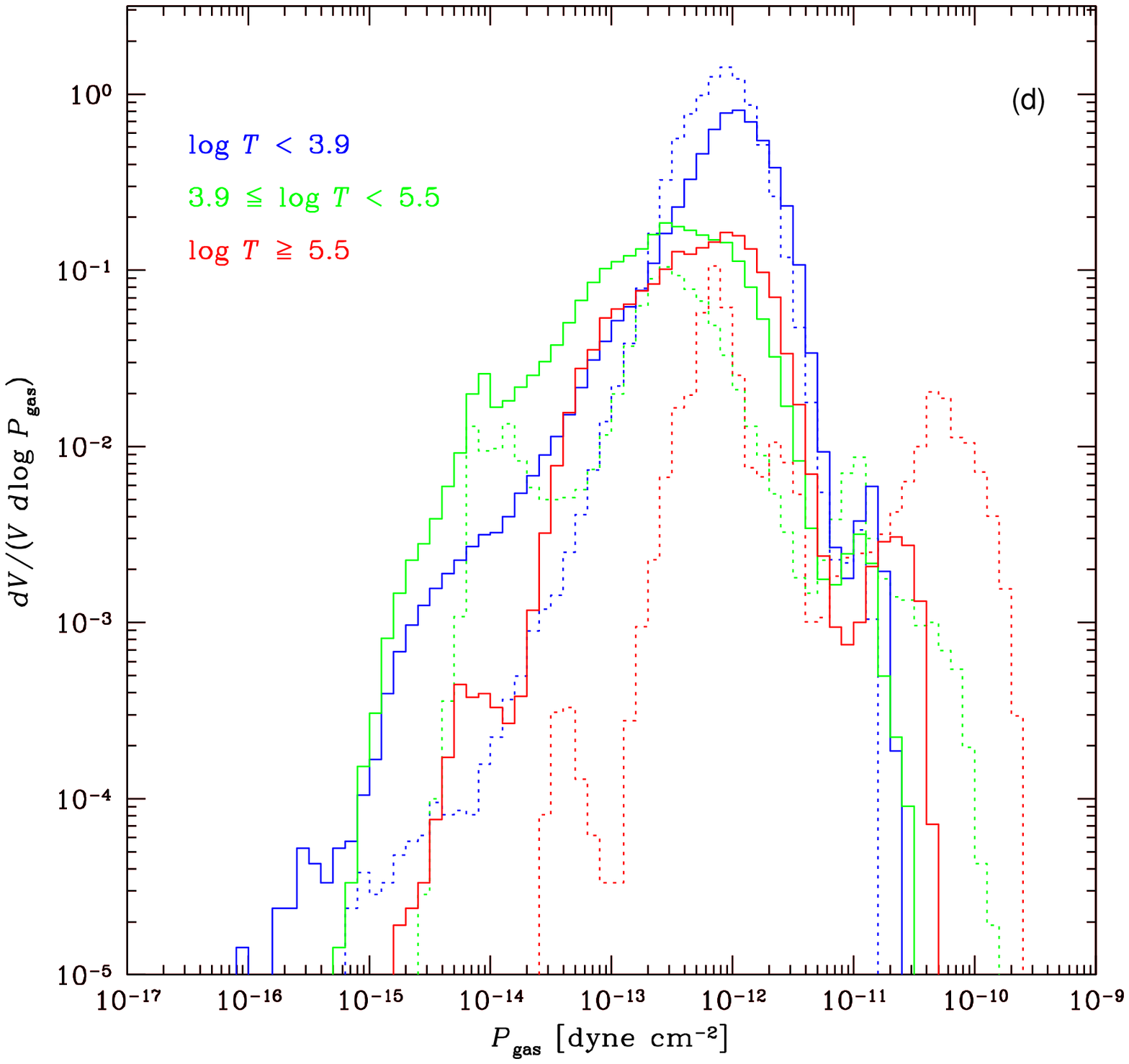,width=0.5\textwidth}
}}
\caption{\label{fig:mhd-pdf}  Volume-weighted
  PDFs of pressure for cool gas with $\log T < 3.9$ ({\em blue}), warm
  gas with $3.9 < \log T < 5.5$ ({\em green}), and hot gas with $\log
  T > 5.5$ ({\em red}) {\em (a)} at different resolutions of 1.56~pc
  ({\em solid}, model M8), and 3.13~pc ({\em dotted}, model M7), at
  25~Myr, {\em (b)} at different times of 15~Myr ({\em dashed}) and
  25~Myr ({\em solid}) in the 1.56~pc resolution model M8, {\em (c)}
  at different SN rates of Galactic ({\em dotted}, model M6) at
  100~Myr, and four times Galactic ({\em solid}, model M8) at 25~Myr,
  and {\em (d)} at different initial magnetic field strengths of
  2~$\mu$G ({\em solid}, model M8) and 5.8~$\mu$G ({\em dotted}, model
  M4) at 25~Myr.  
}
\end{figure}

We have shown that our models display a broad range of pressures.  We can
quantify this by examining the pressure PDFs, as shown in
Figure~\ref{fig:mhd-pdf}. The first point to note, which we will
return to below, is that all of our models show roughly log-normal
pressure PDFs, unlike the power-law distributions predicted by the
analytic theory derived by MO77 (see \S~\ref{sub:snrs}).
The observed distributions rather more resemble the pressure
distributions suggested by PNJ97 and PV98 (see \S~\ref{sub:turb}).
Not only is the total distribution broad, even at the Galactic~SN rate
(Fig.~\ref{fig:mhd-pdf}), but so is the distribution for different
components of the interstellar medium individually.  Furthermore, the
typical or median pressure at the center of the PDF can vary with
temperature, as demonstrated by the PDFs of gas at different
temperature.

Our results do not appear to depend strongly on numerical resolution
as demonstrated in Figure~\ref{fig:mhd-pdf}{\em (a)}, although the
details of the history of each SNR and the total amount of energy
radiated away will certainly depend on the resolution, as well as on
our neglect of the physics of the conductive interfaces between hot
and warm gas.  The pressure PDFs also appear to be stable over time,
as shown by the comparison of multiple times in
Figure~\ref{fig:mhd-pdf}{\em (b)}, except for the hot gas, especially
at the high-pressure end.  Individual young SNRs produce discrete
bumps that move left towards lower pressures as time passes,
eventually merging with the overall distribution.

Figure~\ref{fig:mhd-pdf}{\em (c)} shows that increasing the SN rate
produces a broader distribution of pressures, as generally predicted
by both PNJ97 and PV98, and quantified below.  Finally,
Figure~\ref{fig:mhd-pdf}{\em (d)} shows that increasing the magnetic
field strength reduces the amount of hot gas as SN remnants are better
confined, and reduces the total amount of gas with low thermal
pressure.

\begin{figure}  
\centerline{\hbox{
\psfig{file=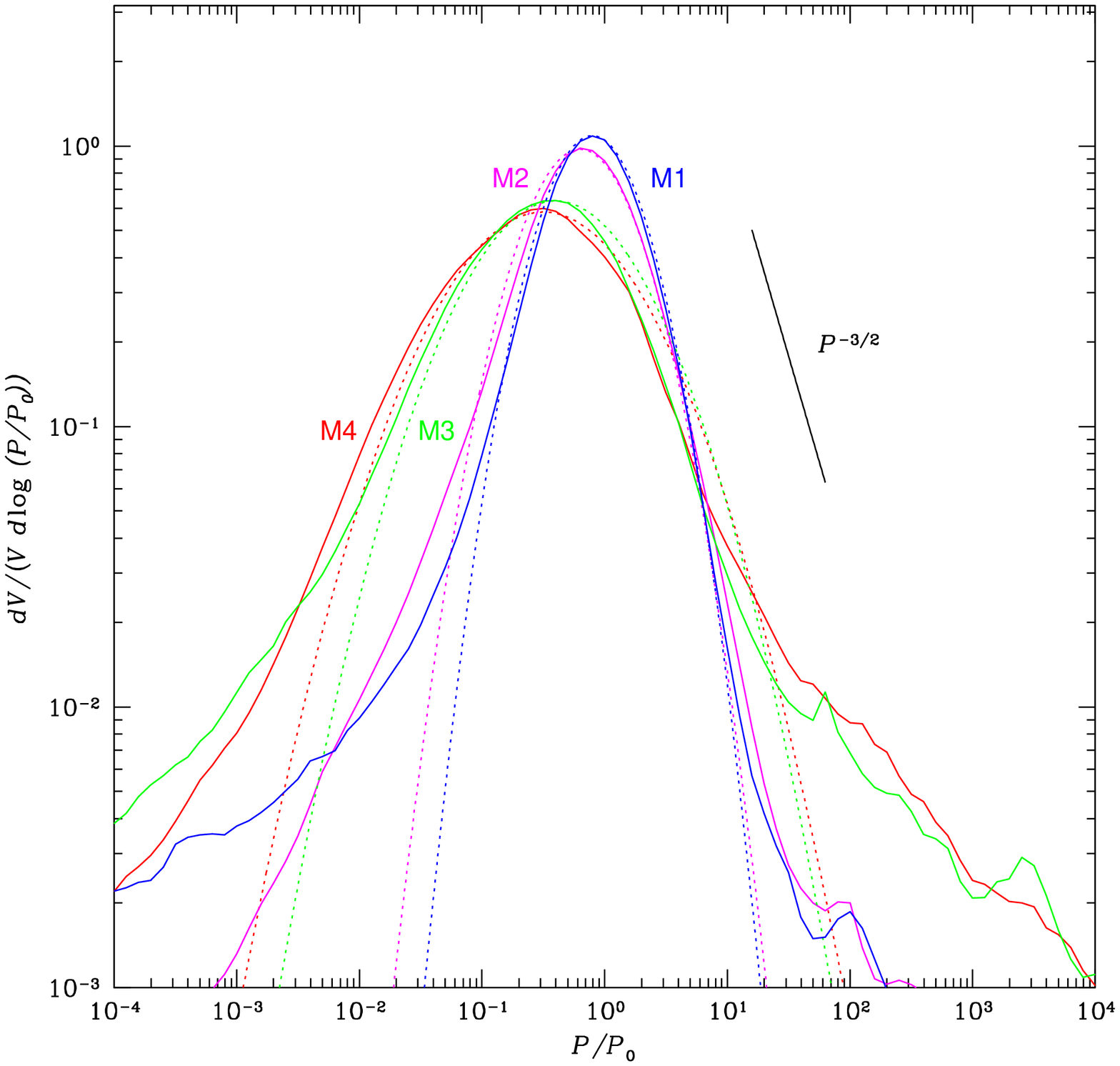,width=0.5\textwidth,angle=0} 
\psfig{file=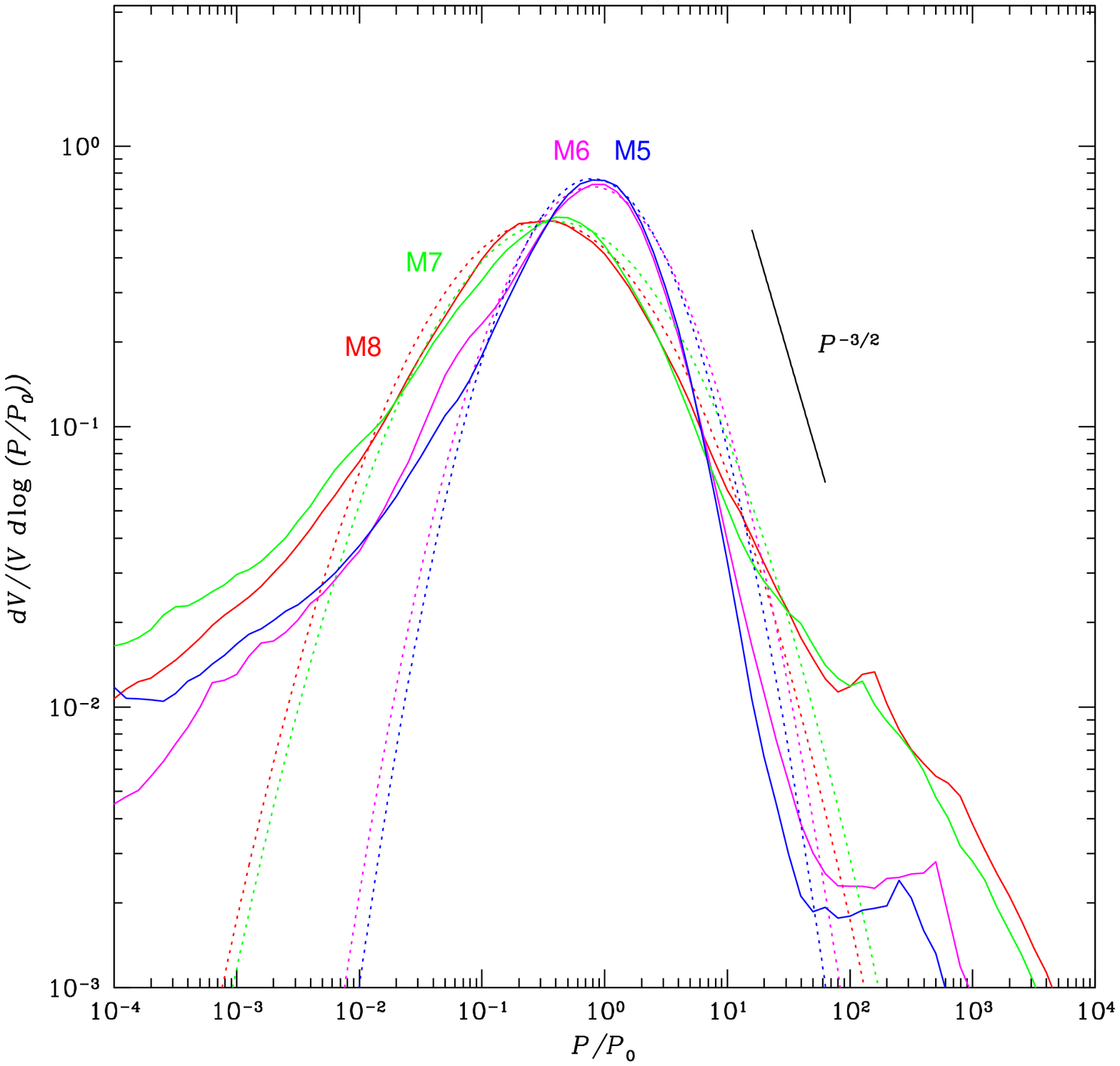,width=0.5\textwidth,angle=0} 
}}
\caption{\label{fig:pdf-rate-av} Volume-weighted PDFs  ({\em
    solid}) and best fit Gaussians ({\em dotted})of thermal pressure
  for magnetic field strengths of {\em (a)} $5.8 \mu$G and {\em (b)}
  $2 \mu$G for models with resolutions of 1.56~pc ({\em red, purple}),
  and 3.13~pc ({\em green, blue}), and SN rates of Galactic ({\em
    blue, purple}), averaged over 50 times from 75--100~Myr, and four
  times Galactic ({\em red, green}), averaged over 50 times from
  18.75--25~Myr.  Solid lines show the prediction of MO77 (taking
  $14/9 \simeq 3/2$).  Model names are noted on the figure.  }
\end{figure}

Using the results for different SN rates illustrated in
Figure~\ref{fig:mhd-pdf}{\em (c)}, we can quantitatively test how well
our results agree with the analytic descriptions of PV98 and PNJ97.
We fit a simple log-Gaussian commensurate with the form of
equation~(\ref{eq:pdfiso}).  These fits are shown in
Figure~\ref{fig:pdf-rate-av} for all of our models.  Although the
Gaussians fit the peaks of our pressure PDFs reasonably well,
exponential tails do appear in almost all cases.  

We further include a line corresponding to the power law predicted by
MO77, as given in our equation~(\ref{mo77}), noting that
Figure~\ref{fig:pdf-rate-av} shows the log of the PDF, reducing the
power law expected by a factor of unity from 23/9 to $14/9 \simeq
3/2$.  This prediction falls in between the values seen for the
Galactic rate and the higher SN rate near the peak, and overpredicts
the high-pressure tail at the Galactic rate, while underpredicting the
tail at the higher rate.  It does not appear to give a general
description of our results. We further note that there is no hint of a
gap in the differential pressure distribution corresponding to the
lack of pressures just below the critical pressure for shell cooling
in MO77 (see discussion above below eqn.\ (\ref{qeqn}).

\begin{figure}[htbp]   
  \centerline{
\psfig{file=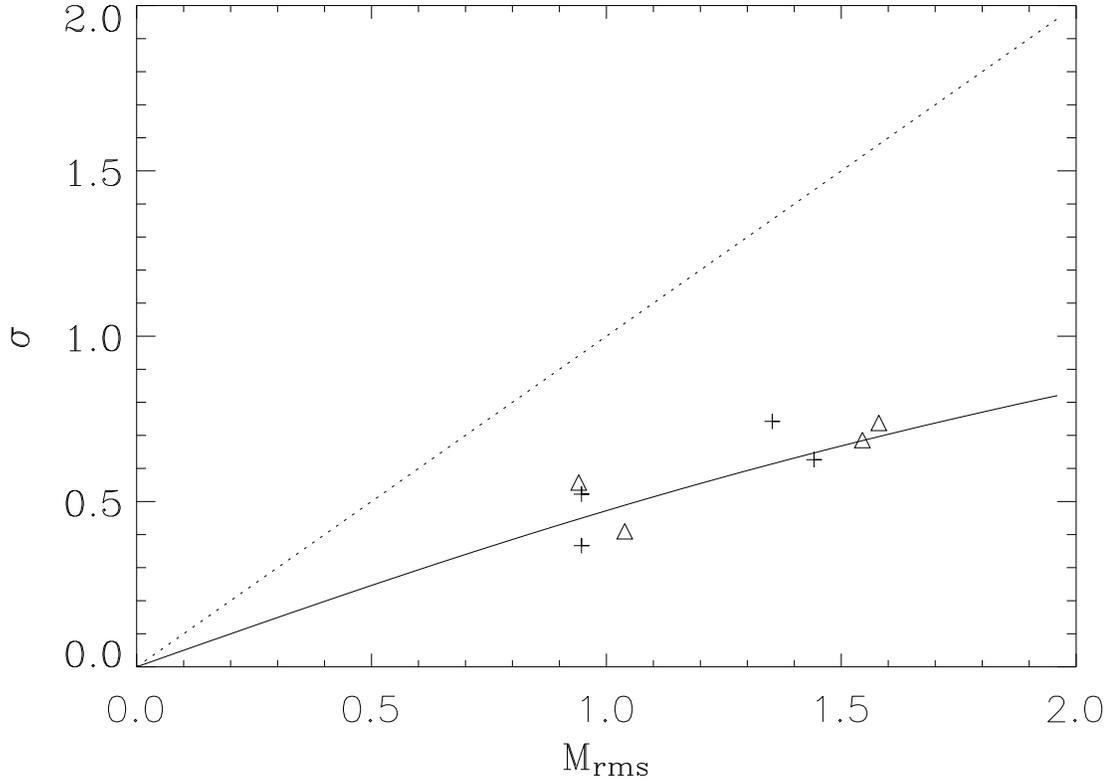,width=\textwidth,angle=0}
}
  \caption{Comparison of predictions of PDF width as a function of
    volume-weighted rms Mach numbers from analytic descriptions of
    supersonic, uniformly-driven, hydrodynamic, isothermal turbulence
    by PNJ97 ({\em solid}) and PV98 ({\em dotted}) to our
    low-resolution ({\em crosses}) and high-resolution ({\em
      triangles}) results. The description of PNJ97 agrees far better
    with our results.  Low-resolution models have widths very similar
    to their corresponding high-resolution models, although their Mach
    numbers differ more significantly in the high SN rate cases. 
}
  \label{fig:an-num-comp}
\end{figure}

We now compare the value of the dispersion of log pressure $\sigma$
that we find in our Gaussian fits to the predictions of PNJ97 and PV98
given in \S~\ref{sub:turb}.  As an approximation to 
$x_0$, the log of the average pressure, 
we take the peaks of the PDFs (the modes).  We then use the rms
velocities of each model, to predict $\sigma_x$.  In
Table~\ref{tab:runs} we give these 
values, and in Figure~\ref{fig:an-num-comp} we compare the fits to the
widths predicted by the rms Mach number for both descriptions.  Our
results agree reasonably well with the prediction for PNJ97, and
disagree strongly with that for PV98.

Ostriker et al.\ (2001) and
Li, Klessen, \& Mac Low (2003) have noted the failure of PV98 to
predict the PDFs for three-dimensional, uniformly-driven turbulence
with both isothermal and non-isothermal equation of state, suggesting
that the difference in results between PNJ97 and PV98 is due to the
use of one-dimensional runs in PV98.  The generally good agreement we find
here with PNJ97 is perhaps somewhat surprising, given that their result was
derived for uniformly-driven, isothermal, turbulence, while we include
a non-isothermal equation of state, supernova explosions, and a
magnetic field.  All of these effects appear to play only a secondary
role, however.  Li et al.\ (2005) show that the
assumption of an isothermal equation of state appears to be sufficient
to explain the gross star-forming properties of galaxies, so perhaps
our result is not entirely far-fetched.

\begin{figure}[htbp]  
\centerline{\hbox{
\psfig{file=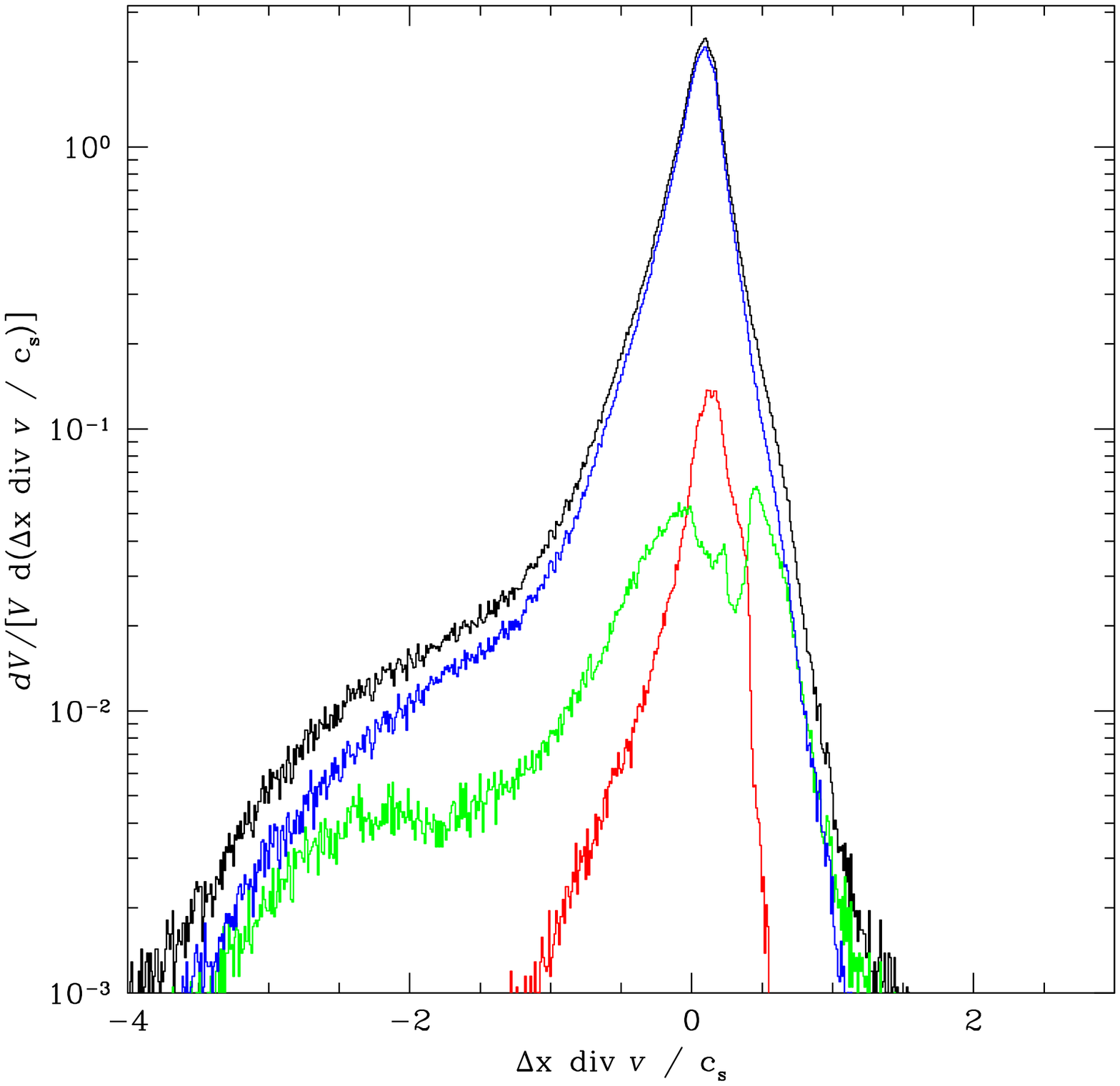,width=0.5\textwidth}
\psfig{file=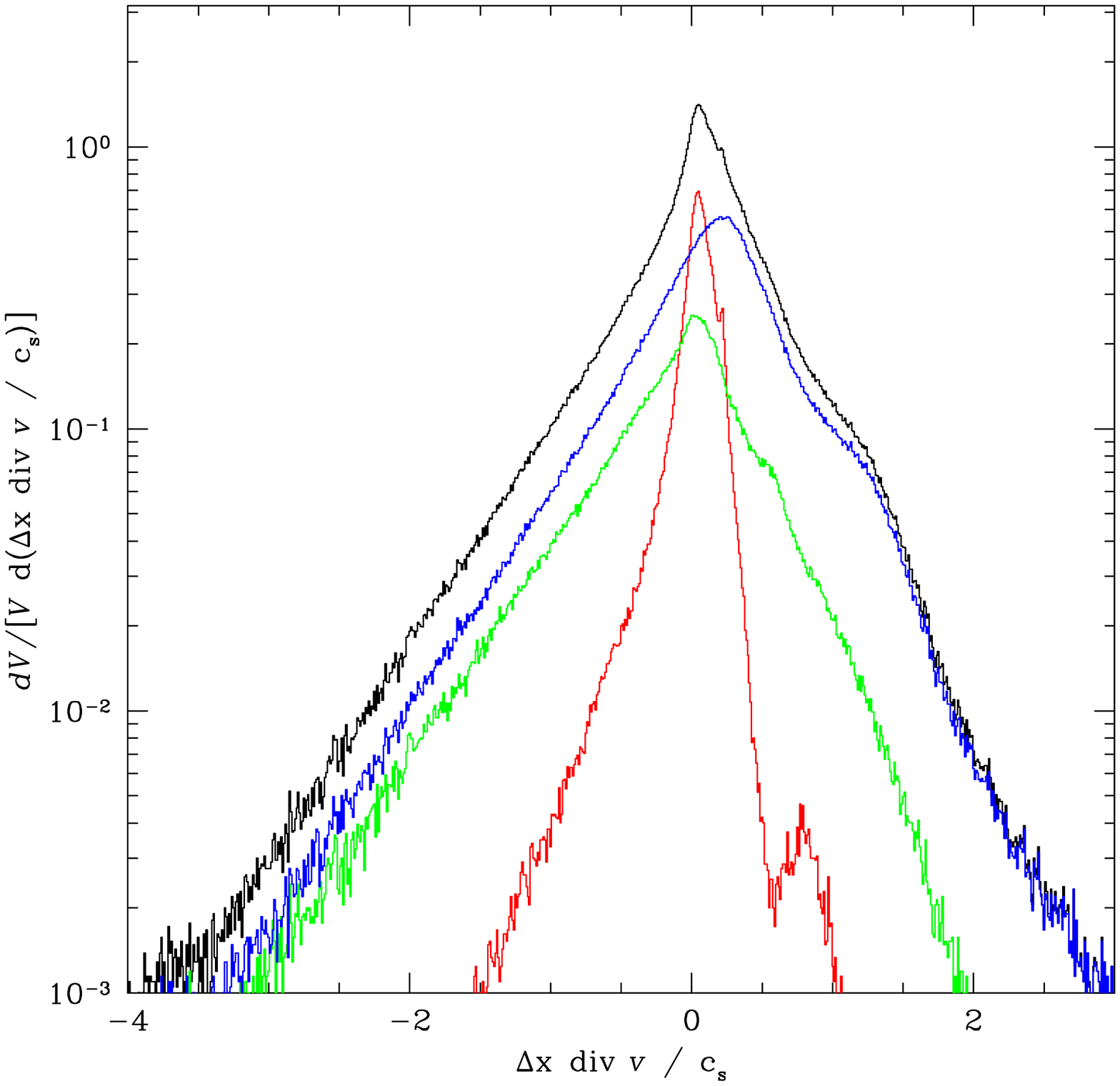,width=0.5\textwidth}
}}
\caption{\label{fig:divv-pdf} Volume-weighted PDF of the divergence of
  velocity normalized by sound speed $\nabla \cdot \vec{v} / c_s$ for
  low field models with supernova rates of {\em (a)} Galactic (model
  M6) and {\em (b)} four times Galactic (model M8). Negative values
  greater trace compressions and shocks, while positive values trace
  rarefaction waves and expanding flows.  Note the general
  symmetry between positive and negative values at the peaks of the
  PDFs, supporting the theory that the density and pressure
  distribution is determined by a random series of shocks and rarefactions.} 
\end{figure}

The Gaussian form of the PDF used by both PNJ97 and PV98 was inspired
by the theory of V\'azquez-Semadeni (1994) who argued that the density
fluctuation spectrum in a strongly supersonic flow is built up by a
random series of compressions and rarefactions.  We can directly trace
compressions and rarefactions by examining the distribution of $\nabla
\cdot \vec{v}$ in our models (Fig.~\ref{fig:divv-pdf}). This
distribution is generally symmetric around the peak at zero, with
compressions on the negative side and rarefactions on the positive
side appearing nearly equally common.  This supports the idea that the
density and pressure fluctuation spectra are generated by repeated
compressions and rarefactions occurring with equal probability.

\begin{figure} 
\centerline{\hbox{
\psfig{file=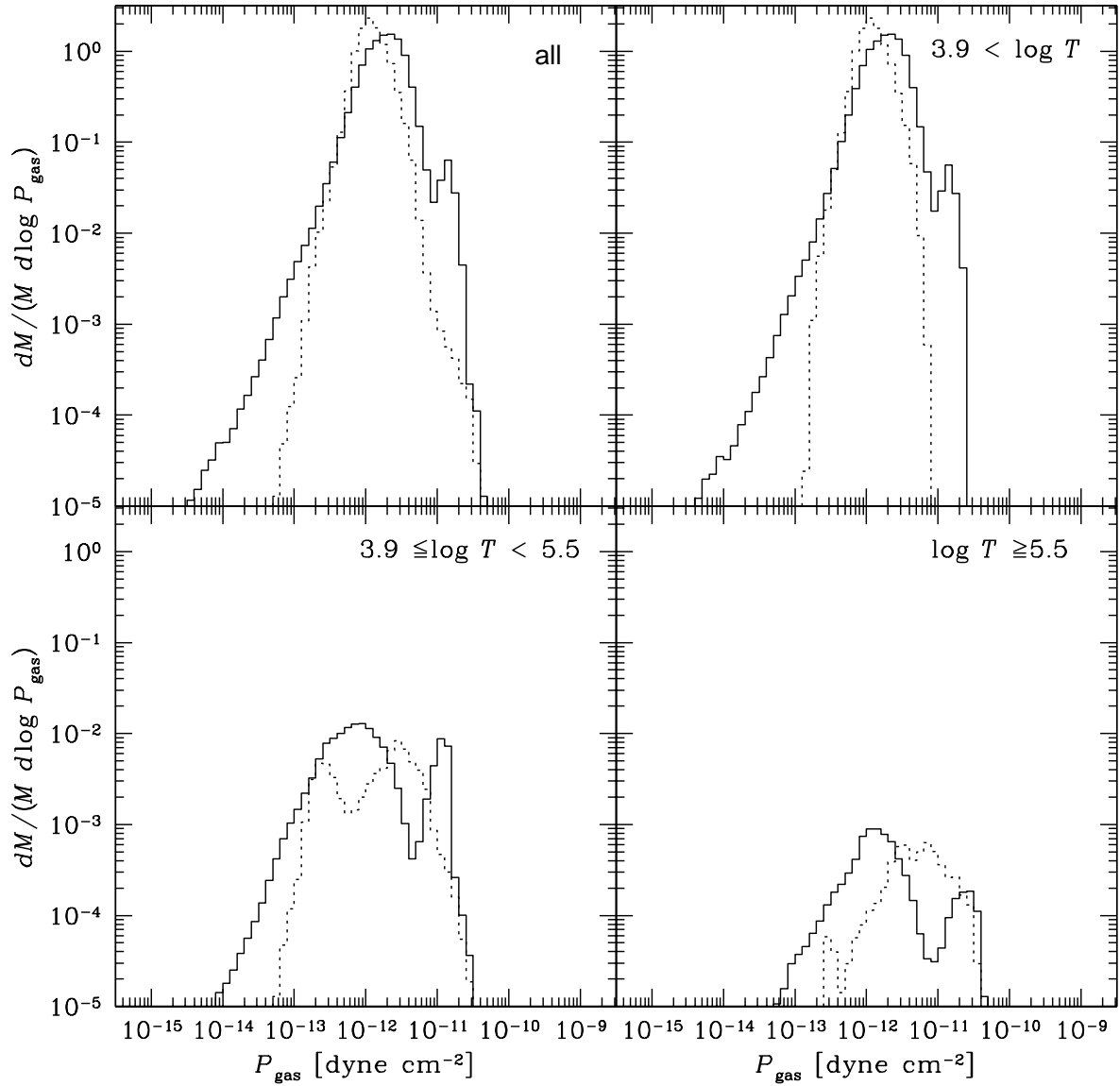,width=\textwidth}
}}
\caption{\label{fig:mass-pdf} Mass-weighted PDF of pressure in 
  low field models with SN rates of Galactic ({\em dotted}, model M6)
  at a time of 100 Myr, and four times Galactic ({\em solid}, model
  M8) at 25 Myr, for all the gas ({\em upper left}), cool gas with
  $\log T < 3.9$ ({\em upper right}), warm gas with $3.9 < \log T <
  5.5$ ({\em lower left}), and hot gas with $\log T > 5.5$ ({\em lower
    right}).  Most of the mass is found in cold gas, with a broad
  distribution around the peak pressure.  }
\end{figure}

The question of how mass is distributed among regions of different
pressure becomes important when considering questions such as the
potential ram-pressure confinement of molecular clouds.  The mass
distribution might be expected to markedly diverge from the volume
distribution given by the PDFs shown previously, as most of the hot gas
resides at very low densities.  In Figure~\ref{fig:mass-pdf} we show
mass-weighted distribution functions from each high-resolution model.
The cold gas dominates the mass-weighted PDF, but it is found at the
same wide range of pressures, with roughly the same peak pressure, as
was suggested by the volume-weighted PDFs shown earlier.  In
particular, a substantial fraction of the mass in the cold gas lies at
pressures five to ten times higher than the average pressure, even in
the absence of self-gravity.

\section{Comparison to Observations}
\label{sec:observations}

\subsection{Ionized and Atomic Gas}

Evidence for a broad distribution of interstellar pressures had
already been noted as early as the work on excited levels of C{\sc i}
by Jenkins \& Shaya (1979) and Jenkins et al.\ (1983).  The point was
sharpened with the direct comparison of nearby pressures by Bowyer et
al.\ (1995), who compared the pressure of the interstellar medium
impinging on the heliosphere with the pressure of the interstellar
medium averaged over a 40~pc line of sight as measured with the {\em
Extreme Ultraviolet Explorer}.  Jenkins \& Tripp (2001)
have used the STIS to extend the work on C{\sc i} with better resolved
data, concluding that the pressure varies by over an order of
magnitude both above and below the average value in a small fraction
of the gas.  Our models naturally explain these variations in the
context of a SN-driven interstellar medium.  

Jenkins et al.\ (1983) compared their results to the analytic theory
of MO77, described above in \S~\ref{sec:an-th}.  They found a
substantially greater column density of low-pressure material than
that predicted, and rather less high-pressure material.  As the
analytic theory could not predict pressures much less than average,
while our models show a broad range of rarefaction waves producing a
log-normal pressure distribution around the mean, the low-pressure
results appear consistent.  Jenkins et al.\ were observing emission
from excited states of C{\sc i} in low-temperature gas.  We find that
much or most of the high-pressure gas resides in the hot medium, so,
although we do predict a greater total volume of high-pressure gas
than was predicted by MO77, we expect that most of it would have been
ionized, and therefore not observable by Jenkins et al.\ (1983),
explaining their low column densities of high-pressure material.

Bowyer et al.\ (1995) and Bergh\"ofer et al.\ (1998) derived a lower
limit to average pressure derived from extreme ultraviolet emission
along lines of sight to H~{\sc i} clouds in the Local Bubble. The
cloud distances were determined using photometry of stars superposed
on the cloud boundaries.  They found a pressure of $P/k >
$13,500--16,500~K~cm$^{-3}$ for the hot medium, with the error
depending mostly on the choice of plasma code used to derive the
emission.  They compared this to the value for the pressure in the
Local Cloud of 700--760~K~cm$^{-3}$ derived by Frisch (1994) from
scattering of solar He~{\sc i}~584\AA radiation from helium flowing in
from the cloud through the heliosphere.  McKee (1996) argues that a
more careful treatment of the unknown ionization fraction could lead
to a local pressure a factor of three higher, reducing, but not
eliminating, the discrepancy.  We find greater than order of magnitude
variations in our models, with pressures reaching values as low as a
few hundred K~cm$^{-3}$ in isolated regions in both sets of models.
Interestingly, the lowest pressures in our models occur at moderately
low densities of order 0.1~cm$^{-3}$, while the density derived for
the Local Cloud by, for example, Quemerais et al.\ (1994) is
0.14~cm$^{-3}$.

An example of a region with high and low pressure regions intertwined
in the manner suggested by the EUV observations is the superbubble
seen in the corners of Figure~\ref{fig:mhd-cut}, particularly the part
in the upper left corner (note that we are using periodic boundary
conditions).  In this region, even local pressure equilibrium is
lacking on scales of tens of parsecs, in contrast to all multi-phase
models.  Similar regions form regularly over time. 

Jenkins \& Tripp (2001) found that their results implied an effective
polytropic index in the cold gas of $\gamma > 0.9$, somewhat higher
than the $\gamma = 0.72$ derived by Wolfire et al.\ (1995) for this
gas.  The suggestion advanced to explain this is that the regions
being compressed may be smaller than the cooling length scale, and so
may begin to behave adiabatically.  An alternative explanation may be
drawn from the broad range of pressures at which gas cools in our
model: the (relatively sudden) pressurization may happen prior to
cooling, rather than to already cooled gas, as required by the
derivation of $\gamma$ by Jenkins \& Tripp.

\subsection{Molecular Gas}

Molecular clouds are observed to have broad linewidths suggesting that
they are subject to pressures as high as $10^5$~K~cm$^{-3}$.  Larson
(1981) was one of the first to suggest that the effective pressure was
due to the self-gravity of the cloud.  Since then, several authors
have found that most of the individual clumps in molecular clouds are
not in hydrostatic balance between turbulent pressure and
self-gravity, but rather are confined by an external pressure (Carr
1987; Loren 1989; Bertoldi \& McKee 1992).  The explanation offered
for this by Bertoldi \& McKee (1992) was that the entire cloud was
still subject to self-gravity, even though individual clumps were not,
and so the effects of self-gravity on large scales produced pressures
that confined the clumps.

In our models, we find pressures in high-density regions of order
$10^5$~K~cm$^{-3}$ in the absence of self-gravity, as shown in
Figure~\ref{fig:mhd-dscat}.  These pressures are sufficient to confine
observed clumps without invoking self-gravity, suggesting that
observed molecular clouds may be primarily pressurized by the ram
pressure of the turbulent flows in which they are embedded rather than
being self-gravitating objects.  Simulated observations of turbulent
flows suggest that mass-linewidth relations thought to indicate that
they are in virial equilibrium may actually be due to a combination of
the properties of the turbulence itself in the case of the
size-linewidth relation, and the properties of the observations in the
case of the size-mass relation (V\'azquez-Semadeni,
Ballesteros-Paredes, \& Rodriguez 1997; Ballesteros-Paredes, \& Mac
Low 2002).

Ram pressure is a double-edged sword, however, that can destroy clouds
as easily as creating them.  This is consistent with the suggestion
that they are transient objects with lifetimes of under $10^7$~yr
first made by Larson (1981), and emphasized by Ballesteros-Paredes, et
al.\ (1999a), Elmegreen (2000, 2002), and Hartmann et al.\ (2001).
Relying on ram pressures rather than self-gravity to confine observed
clumps in molecular clouds would also be consistent with the results
of simulations of hydrodynamical and MHD driven, self-gravitating,
isothermal turbulence that showed self-gravity only acting on small
scales, with turbulent flows dominating the large scales (Klessen et
al.\ 1999; Heitsch et al.\ 2001).  The same flows that confine and
destroy the clumps also drive the turbulence observed within them, as
the background flow stretches, twists, forms, and destroys dense
regions contained within it.

Our prescription of isochoric nonequilibrium cooling for $10^4 \mbox{
  K} < T < 10^6$~K, and more especially the absence of the physically
expected region of thermal instability at $T< 10^4$~K prevents us from
having any degree of confidence in the exact amounts of cold gas
produced at any given pressure.  However, our qualitative results
appear quite robust, so we expect future work to refine the details
rather than substantially change the picture of molecular clouds
forming in transient, high-pressure, high-density regions produced by
a supersonic, turbulent flow.

\section{Discussion and Summary}
\label{sec:summary}

We have examined the distribution of pressures predicted from
three-dimensional simulations of a SN-driven, magnetized ISM,
neglecting vertical stratification and self-gravity, but including a
distributed heating function.  In all the simulations we have run, we
find pressures distributed over rather more than an order of magnitude
around the mean.  The pressure PDF can be fit well with a log-normal
having width given by the analytic description of PNJ97 based on
three-dimensional simulations of isothermal turbulence, and
disagreeing with the one-dimensional results of PV98.  The log-normal
distribution is predicted by the theory of V\'azquez-Semadeni (1994)
that density fluctuations in a supersonic turbulent flow result from a
sequence of random compressions and rarefactions. We find that
compressions and rarefactions are indeed symmetrically distributed,
lending further support to this idea.  The log normal form differs
from the power-law distribution predicted by MO77.
The only isobaric regions found in our models are the interiors of
young SNRs; these are also the only laminar flow regions in an
otherwise turbulent environment.

Most of the mass is in the turbulent gas, however, as is most of the
volume, even in models driven with four times the Galactic supernova
rate.  This result is consistent with the results of Korpi et al.\ 
(1999), Avillez (2000), and Avillez \& Breitschwerdt (2004).  The last
group performed a resolution study to establish that numerical
diffusion was not causing hot gas to cool unphysically.  We will study
the filling factors of our models in more detail in future work.

We find a broad range of pressures, and a substantial fraction of
associated densities, far from the thermal equilibrium values.  This
limits the predictive usefulness of phase diagrams based on thermal
equilibrium, although thermal equilibrium at the local pressure is
still the mildly favored state.  Gas pressures appear to be determined
dynamically.  Each individual parcel of gas seeks local thermal
equilibrium at the pressure imposed on it by the turbulent flow. As
suggested by previous authors, the phase diagram is only locally
valid.  Isobaric thermal instability in such an environment will lead
to regions of the phase diagram being mildly disfavored, but no more.
V\'azquez-Semadeni et al.\ (2000) and Gazol et al.\ (2001) used
two-dimensional simulations with more physically-motivated cooling
curves at lower temperatures to conclude that the effects of the
thermal instability will be barely visible in the overall probability
distribution functions (but note that these simulations did not
resolve cooling regions well, so the results should be taken with
caution).

Our results appear consistent with observations that have repeatedly
shown the ISM not to be isobaric, including those by Jenkins \& Shaya
(1979), Jenkins, Jura \& Loewenstein (1983), Bowyer et al.\ (1995),
Bergh\"ofer et al.\ (1998), and Jenkins \& Tripp (2001).  Although
heating and cooling rates remain important for determining the local
density and temperature, they do not produce a global multi-phase
medium with well-separated regions of different temperature gas
because of the wide range of pressures present and the dynamical
processing of the gas.  Rather, a more continuous distribution of
temperatures and densities is always present.

The inference that most molecular clouds must be gravitationally bound
because of their high observed confinement pressures are called into
question by our results.  Regions with densities approaching the
overall densities of GMCs, and pressures an order of magnitude above
the average interstellar pressure appear in our simulations even in
the absence of self-gravity.  This supports suggestions that
star-forming molecular clouds may be transient, turbulently-driven,
objects (Ballesteros-Paredes et al.\ 1999ab, Elmegreen 2000, 2002, Hartmann
et al.\ 2001). Objects that do gravitationally collapse from large
scales as described, for example by Kim \& Ostriker (2001), Elmegreen
(2002), or Li et al.\ (2005) will also form molecular
gas, but will quickly form starburst knots in a burst of violent,
unimpeded star formation.

Magnetic pressures are also broadly distributed, and do not correlate
well with thermal pressures.  The ratio of thermal to magnetic
pressure $\beta$ takes on values ranging from 0.1 to as much as 100 in
gas with temperatures under $10^4$~K, and much higher values in hot
gas. The ratio of fluctuating to mean field in our models $\langle
\delta B\rangle / \langle B \rangle=0.5$--1, high enough to suggest
that field strengths derived from observations of polarized synchrotron
emission may be overestimated (Beck 2001).

\acknowledgments We thank J. Ballesteros-Paredes, D. Cox, E. B.
Jenkins, C. F. McKee, T. Passot and E. V\'azquez-Semadeni for valuable
discussions, clarifications of their work cited here, and useful
comments on drafts.  Two anonymous referees of an earlier version of
this paper (Mac Low et al.\ 2001) also made a number of useful
suggestions, including suggesting the comparision to PNJ97.  M-MML and
MAA acknowledge support from NSF CAREER grant AST 99-85392, while DSB
acknowledges NSF grants CISE 1-5-29014, DMS 02-04640, and AST
00-98697. JK acknowledges support from the Astrophysical Research
Center for the Structure and Evolution of the Cosmos funded by the
Korea Science and Engineering Foundation. Computations presented here
were performed at the National Center for Supercomputing Applications,
which is supported by the NSF, and on the Linux cluster at the Korea
Astronomy Observatory.  This research has made use of NASA's
Astrophysics Data System Abstract Service.

\clearpage
\begin{center} {\Large Figures} \end{center}

\clearpage
\begin{deluxetable}{lcccccccccc}
\tablecaption{Model Properties. \label{tab:runs}}
\tablewidth{0pt}
\tablecolumns{11}
\tablehead{
	 \colhead{model}
	&\colhead{$\tau_{\rm SN}$\tablenotemark{a}}
        &\colhead{$\Delta x_{\rm min}$} 
        &\colhead{$B_0$}  
        &\colhead{$\langle B\rangle$}  
        &\colhead{$\langle \delta B \rangle$}  
        &\colhead{\underbar{$\langle \delta B \rangle$}}  
        &\colhead{$M_{\rm rms}$\tablenotemark{b}}
        &\colhead{$\sigma$\tablenotemark{c}}
\\
         \colhead{}    
        &\colhead{}      
	&\colhead{(pc)}
	&\colhead{($\mu$G)}
	&\colhead{($\mu$G)}
	&\colhead{($\mu$G)}
        &\colhead{$\langle B \rangle$}
        &\colhead{}
        &\colhead{}
}
\startdata
M1  & 1 & 3.13 & 5.8 & 6.258 & 3.063 & 0.489 & 0.947 & 0.3666 \nl
M2  & 1 & 1.56 & 5.8 & 6.443 & 3.470 & 0.539 & 1.039 & 0.4100 \nl
M3  & 4 & 3.13 & 5.8 & 6.685 & 4.430 & 0.663 & 1.442 & 0.6266 \nl
M4  & 4 & 1.56 & 5.8 & 7.049 & 4.778 & 0.678 & 1.545 & 0.6859 \nl
M5  & 1 & 3.13 & 2.0 & 2.436 & 1.854 & 0.761 & 0.947 & 0.5222 \nl
M6  & 1 & 1.56 & 2.0 & 2.589 & 2.071 & 0.800 & 0.941 & 0.5575 \nl
M7  & 4 & 3.13 & 2.0 & 2.587 & 2.610 & 1.009 & 1.353 & 0.7423 \nl
M8  & 4 & 1.56 & 2.0 & 3.071 & 3.067 & 0.999 & 1.580 & 0.7378 \nl
\enddata
\tablenotetext{a}{SN rate in terms of the Galactic SN rate}
\tablenotetext{b}{Volume-weighted rms Mach number}
\tablenotetext{c}{Dispersions of log of pressure derived from fits to
  PDFs} 
\end{deluxetable}

\begin{deluxetable}{lc}
\tablecaption{Cooling Times \label{tab:cool}}
\tablewidth{0pt}
\tablecolumns{2}
\tablehead{
\colhead{$\log_{10} T$} & \colhead{$t_{\rm cool} (n/1\mbox{
cm}^{-3})^{-2}$}  \\
\colhead{(K)} & \colhead{(yr)}
}
\startdata
3  &   1.7(5) \nl
4  &   6.0(3) \nl 
5  &   1.6(3) \nl
6  &   4.4(4) \nl
7  &   1.2(6) \nl
\enddata
\end{deluxetable}

\end{document}